\documentclass[sn-apa]{sn-jnl}

\usepackage{color,soul}
\usepackage{natbib}
\usepackage{graphicx}
\usepackage{amsmath,amssymb}
\usepackage{rotating}
\usepackage{multirow}
\usepackage{booktabs}
\usepackage{longtable}
\usepackage{appendix}
\usepackage[utf8]{inputenc}
\usepackage[english,vietnamese]{babel}
\usepackage{CJKutf8} 
\usepackage{float}

\usepackage{array}
\newcolumntype{L}[1]{>{\raggedright\let\newline\\\arraybackslash\hspace{0pt}}m{#1}}
\newcolumntype{C}[1]{>{\centering\let\newline\\\arraybackslash\hspace{0pt}}m{#1}}
\newcolumntype{R}[1]{>{\raggedleft\let\newline\\\arraybackslash\hspace{0pt}}m{#1}}

\newcommand{\ie}{\textit{i}.\textit{e}., }
\newcommand{\eg}{\textit{e}.\textit{g}. }
\newcommand{\highlight}[1]{#1}

\jyear{2022}%

\theoremstyle{thmstyleone}%
\theoremstyle{thmstyletwo}%

\theoremstyle{thmstylethree}%

\begin{document}

\title[Attentive DNNs for Legal Document Retrieval]{Attentive Deep Neural Networks for Legal Document Retrieval}


 \author*[1,3]{\fnm{Ha-Thanh} \sur{Nguyen}}\email{nguyenhathanh@jaist.ac.jp}
 \equalcont{These authors contributed equally to this work.}

 \author[2]{\fnm{Manh-Kien} \sur{Phi}}\email{kienpm2205@gmail.com}
 \equalcont{These authors contributed equally to this work.}

\author[2]{\fnm{Xuan-Bach} \sur{Ngo}}\email{bachnx@ptit.edu.vn}

\author[1]{\fnm{Vu} \sur{Tran}}\email{vu.tran@jaist.ac.jp}

\author[1]{\fnm{Le-Minh} \sur{Nguyen}}\email{nguyenml@jaist.ac.jp}

 \author[2]{\fnm{Minh-Phuong} \sur{Tu}}\email{phuongtm@ptit.edu.vn}

\affil*[1]{\orgdiv{School of Information Science}, \orgname{Japan Advanced Institute of Science and Technology}, \orgaddress{\city{Nomi}, \state{Ishikawa}, \country{Japan}}}

\affil[2]{\orgdiv{Department of Computer Science}, \orgname{Posts and Telecommunications Institute of Technology}, \orgaddress{\city{Hanoi}, \country{Vietnam}}}

\affil[3]{\orgdiv{Principles of Informatics Research Division}, \orgname{National Institute of Informatics}, \orgaddress{\city{Tokyo}, \country{Japan}}}

\selectlanguage{english}
\abstract{
Legal text retrieval serves as a key component in a wide range of legal text processing tasks such as legal question answering, legal case entailment, and statute law retrieval. 
The performance of legal text retrieval depends, to a large extent, on the representation of text, both query and legal documents. 
Based on good representations, a legal text retrieval  model can effectively match the query to its relevant documents. 
Because legal documents often contain long articles and only some parts are relevant to queries, it is quite a challenge for existing models to represent such documents.
In this paper, we study the use of attentive neural network-based text representation for statute law document  retrieval. 
We propose a general approach using deep neural networks with attention mechanisms. Based on it, we develop two hierarchical architectures with sparse attention to represent long sentences and articles, and we name them Attentive CNN and Paraformer. 
The methods are evaluated on datasets of different sizes and characteristics in English, Japanese, and Vietnamese. 
Experimental results show that: i) Attentive neural methods substantially outperform non-neural methods in terms of retrieval performance across datasets and languages; 
ii) Pretrained transformer-based models achieve better accuracy on small datasets at the cost of high computational complexity while lighter weight Attentive CNN achieves better accuracy on large datasets; and 
iii) Our proposed Paraformer outperforms state-of-the-art methods on COLIEE dataset, achieving the highest recall and F2 scores in the top-N retrieval task\footnote{This paper is an improved and extended work of \citet{kien-etal-2020-answering}}.
}

\keywords{Legal text retrieval, deep neural networks, hierarchical representation, global attention}



\maketitle

\section{Introduction}\label{sec:intro}

Social relations arise, develop and change daily, so legal documents also need to be promulgated to keep up with the changes of life.
There is apparently an increment in the number of legal cases as well as the number of legal documents in different nations.
In 2020, the number of civil and criminal cases in the US reached more than 500 thousands\footnote{https://www.uscourts.gov/statistics-reports/judicial-business-2020}.
As a civil-law nation, Vietnam has more than 20 types of legal documents with thousands of new documents being issued every week\footnote{https://thuvienphapluat.vn/van-ban-moi}.
From the above situation, it can be seen that the use of automatic systems in finding and retrieving documents that match the needs of users is a mandatory requirement. Because of the importance of correctness in the legal field, the performance of these systems is an important attribute to bring them into real life.
In this paper, we propose an effective legal retrieval approach for statute law using novel architectures of attentive deep neural networks.


For a legal retrieval system, given a query $q$, and a legal corpus $\mathcal{L}$, the system needs to return a set of articles $\mathcal{A}\subseteq\mathcal{L}$ that:
$$
Relevance(q,\alpha) \forall \alpha \in \mathcal{A}
$$

\noindent In which $Relevance$ is a boolean function that indicates if an article is relevant to the given query.

To define the problem without ambiguity, we first need to clarify the concept of relevance.
Dealing with problems in the legal domain requires expert knowledge and understanding in this field.
Information retrieval in this field does not simply mean finding all the texts with the most lexical overlapping with the query.
A good system also needs to consider the meaning of the query as well as the articles to make reliable alignment between them \citep{vsavelka2021legal}.
A relevant article is the one that can be used to answer or validate the lawfulness of a query.
Moreover, each article also needs to be interpreted in the appropriate meaning for a specific given query.
In turn, queries with non-legal vocabulary also need to be mapped to the corresponding knowledge area in the legal domain.

\begin{figure}
  \centering
\includegraphics[width=.9\linewidth]{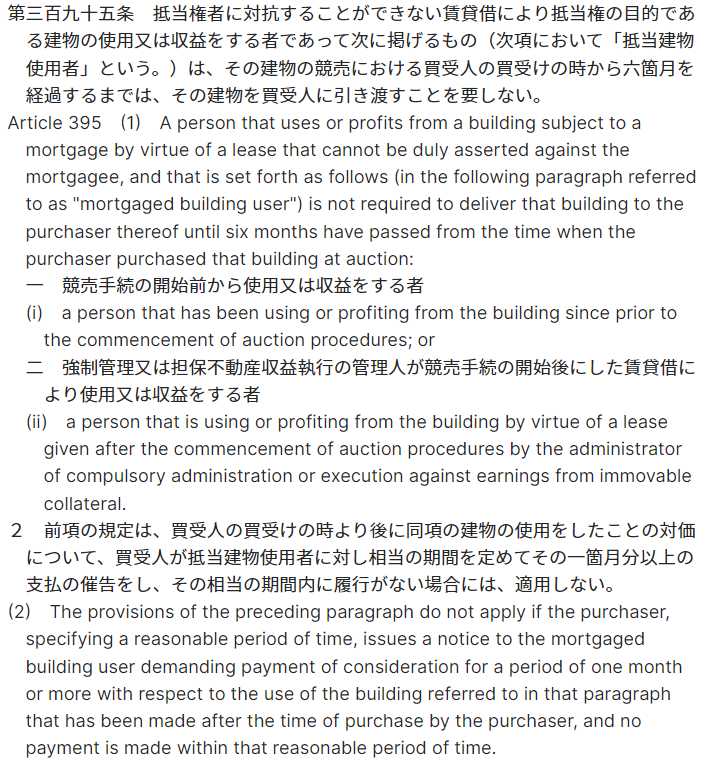}
  \caption{Article 395 in Japanese Civil Code. }
  \label{fig:article_395}
\end{figure}

\begin{figure}
  \centering
\includegraphics[width=.9\linewidth]{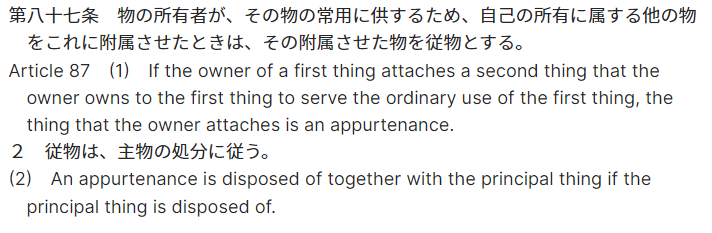}
  \caption{Article 87 in Japanese Civil Code. }
  \label{fig:article_87}
\end{figure}

Merely relying on lexical matching may not be the sufficient approach for this problem. 
For example, with the purpose of confirming the lawfulness of the query \textit{``Extended parts of the building shall be regarded as appurtenance.''}, according to lexical matching result, Article 395 in the Japanese Civil Code (Figure~\ref{fig:article_395}) is the best candidate. 
This article contains many words in common with the given query.
However, the most important word \textit{``appurtenance''} does not appear in Article 395.
The correct article to answer this query is Article 87 (Figure~\ref{fig:article_87}), a shorter article that contains fewer words in common with the given query.
This article does not mention any \textit{``building''} in its content but can be used to verify the lawfulness of this query.
Hence, the better the system understands the semantics of the concepts, the better the performance it can obtain.
Building an accurate legal document retrieval system, therefore, depends heavily on good text representation methods. 

Recently, deep neural network models are very successful in text representation in a wide range of tasks. 
In their development, there are various architectures proposed such as convolutional neural networks (CNNs) \citep{kim2014convolutional,shen2014latent,severyn2015learning,vaswani2017attention}, recurrent neural networks (RNNs) \citep{mikolov2011extensions}, LSTMs \citep{wang2016attention,palangi2016deep,mueller2016siamese,chen2017enhanced,Bach:2019,Bach:2019b} and gated recurrent units (GRUs) \citep{tang2015document}.
Most notably, Transformers \citep{vaswani2017attention} leveraging attention mechanism becomes a well-known approach, its pretrained variants like BERT \citep{devlin2018bert}, BART \citep{lewis2019bart}, GPTs \citep{radford2018improving,radford2019language,brown2020language} achieve impressive results in a wide range of natural language processing tasks. 

\highlight{
Although there are differences among legal systems, they can be classified and generalized into two main theoretical constructs, common law and civil law \citep{husa2016future}.
In the context of civil law tradition, the legal retrieval problem can be done at the document level, article level, or sentence level. 
Through surveying legal consulting activities in civil law nations like Japan, Germany and Vietnam, we found that retrieval at the article level is a popular approach to answer a legal question. 
This survey was conducted through consultation with law professors, attorneys, and investigating scholarly materials \citep{shao2020bert,rabelo2019summary,yoshioka2018overview,nguyen2017knowledge,thanh2021summary} and legal consultant websites in civil law nations like Vietnam\footnote{https://thuvienphapluat.vn}, Japan\footnote{https://keiji.vbest.jp} and Germany\footnote{https://www.anwalt.de}.
In a real situation of legal question answering, the legal consultant often refers to a specific article, neither a whole document nor only a single sentence. 
}
From the technical viewpoint, article-level retrieval has its own challenges.
As can be seen in Table~\ref{tab:example} which demonstrates a legal retrieval-based question answering example, just a few sentences in an answer article contain the necessary information to answer the question. 
This observation inspires us to design an architecture using an attention mechanism to focus on the necessary part of an article for a more effective retrieval system.


In this paper, we focus on the task of retrieving legal documents at the article level, which serves for question answering in civil law systems.
We study on exploiting deep neural networks with attention mechanisms to solve the task. For attention mechanisms, we investigate two recent advanced architectures, i.e., attentive CNNs and self-attention with Transformer, which achieved state-of-the-art results on many NLP tasks. Our contributions can be summarized in the following points:
\begin{enumerate}
    \item We design a general framework for legal document retrieval using deep neural networks with attention mechanisms. Based on this framework, we develop two attentive deep learning models: Attentive CNN and Paraformer, where the latter represents legal \textbf{para}graphs using Trans\textbf{former}. Our approach allows encoding long text by letting the model focus on only the important parts of the text. Compared to previous works, we model legal articles as a hierarchical structure to encode them into the vector space. 
    \item We introduce a Vietnamese dataset for the task, which is much larger than the existing ones. Our dataset is crucial to verify the effectiveness of retrieval models in different languages as well as compare the models’ behavior in different corpus sizes. The dataset is also a good resource for the research community in related problems.
    \item We conduct an empirical study on proposed models using three datasets, including our Vietnamese dataset, and the English and Japanese datasets from COLIEE\footnote{https://sites.ualberta.ca/~rabelo/COLIEE2021/}. Experimental results show that our models outperform existing methods, both non-deep learning and deep learning ones. Although both Attentive CNN and Paraformer are effective for the task, each model is superior to the other in specific situations. Our results also indicate that using transformer-based pre-trained models can improve the performance of retrieval models, especially when we only have a relatively small training dataset. 
\end{enumerate}

The rest of this paper is structured as follows. Section 2 describes related work. Section 3 presents three datasets used in our experiments, i.e., Vietnamese, English, and Japanese. In Section 4, we introduce our general framework for legal text retrieval and two retrieval models. Experimental results and discussions are described in Section 5. Finally, Section 6 concludes the paper and discusses future work.  



\section{Related Work}
Before the application of neural networks became widespread, there were approaches in classical NLP to solve information retrieval tasks \citep{cooper1971definition,luhn1957statistical,salton1988term}.
These methods are mainly based on different lexical matching techniques.
These authors propose logical models as well as statistical models to calculate the similarity between queries and candidates.
The methods have their own advantages such as fast computation speed and applicability to many problems.
Non-neural methods, however, mainly rely on morphology in the text to make decisions.
In natural languages, morphological similarity does not guarantee semantic similarity, so it is difficult to guarantee correctness in semantic similarity using these approaches.
Therefore, these approaches have limited performance in the case that the document-query pairs contain many overlapped texts but no relation in the semantic aspect. 

The legal language can be translated into logical language~\citep{kowalski2021logical}. 
One of the most well-known systems using logical models to perform legal retrieval and reasoning for statute law is PROLEG (PROlog based LEGal reasoning support system)~\citep{satoh2010proleg}. 
This system is empowered by the Japanese Presupposed Ultimate Fact Theory~\citep{ito2008lecture}. 
PROLEG is based on the idea of the \textit{burden of proof} (\ie if a fact is failed to be proved as true, it is considered as false). The relevant rules of the reasoning process can be called out automatically to make reasoning for a query. 
This system, however, requires the queries and legal documents to be formatted in a logical form.
For that reason, the system is not suitable for lay users.

%
Overcoming the challenge of the semantic morphology difference and the burden of logical representation, several neural approaches in information retrieval in both the general domain and legal domain are proposed \citep{palangi2016deep,shen2014latent,huang2013learning,vsavelka2021legal,nguyen2018recurrent}. 
Most of the systems use classical neural network architecture like CNN or LSTM to handle the task. 

For legal text, ~\citet{sugathadasa2018legal} and \citet{tran2020encoded} propose to use neural networks and achieve impressive results.
The authors observe the structure of the legal documents and base on their characteristics to propose novel representation methods.
Through their experimental results, the author demonstrates that their proposals effectively work for the legal domain.
\citet{kien-etal-2020-answering} introduce the neural network architecture that combines CNN and attention mechanisms.
With a lightweight design, our model achieves state-of-the-art results on the Vietnamese legal question-answering dataset.
These works also reveal that the combination between the semantic vectors and the lexical features can boost the overall performance of the systems.

Pretrained neural approaches construct the models in two phases.
In the pretraining phase, the models are trained with general tasks to abstract the relationships between units in the sentences.
After that, the models are finetuned with the specifically designed tasks.
This family of approaches has been demonstrated to be effective in a wide range of natural language processing as well as legal document processing.

The earliest form of pretrained models is the pretrained word embeddings  (Word2Vec~\citep{mikolov2013distributed}, GloVe~\citep{pennington2014glove} or FastText~\citep{mikolov2018advances}). 
With these pretrained embeddings, we can easily find the semantic relationship between words (\eg verify the equation $king=queen+man-woman$).
In the legal domain, authors of Law2Vec \citep{chalkidis2019deep} introduce a variant of word embedding trained on legal corpus and demonstrate its effectiveness.
Recently, pretrained models based on Transformer architecture \citep{vaswani2017attention} achieve state-of-the-art results on many benchmark data, both in the general domain \citep{devlin2018bert,lewis2019bart,radford2018improving,radford2019language,reimers2019sentence,brown2020language} and in the legal domain \citep{yilmaz2019applying,nguyen2020jnlp,yoshioka2021bert,nguyen2021paralaw}.
Pretrained approaches are useful in the case that the training data is limited in quantity.

\section{Datasets}
\label{sec:dataset}
To test the proposed approach, we conduct the experiments on the datasets in three languages: Vietnamese, Japanese, and English.
The Japanese and English datasets are the different versions of the dataset provided by COLIEE.

To build this Vietnamese dataset, we crawled the raw legal documents from the official legal websites \footnote{http://vbpl.vn/tw/pages/home.aspx}\footnote{https://thuvienphapluat.vn} and the queries from the legal consulting websites  \footnote{https://hdpl.moj.gov.vn/Pages/home.aspx}\footnote{http://hethongphapluat.com/hoi-dap-phap-luat.html}\footnote{https://hoidapphapluat.net}. 
The raw data to build the corpus of Vietnamese legal documents contains multiple versions of each law and regulation.
We removed the redundant old versions and remapped the new relevant articles with the corresponding query in the question-answering dataset.
To obtain a good question-answering dataset, we corrected spelling, formatting, grammar errors and filtered out the contents which are confusing, uninformative, or low quality.
The process of reviewing and editing was done with the support of lawyers.
The final version contains 8,586 documents (117,545 articles) and 5,922 legal queries.

\begin{figure}
  \centering
\includegraphics[width=.9\linewidth]{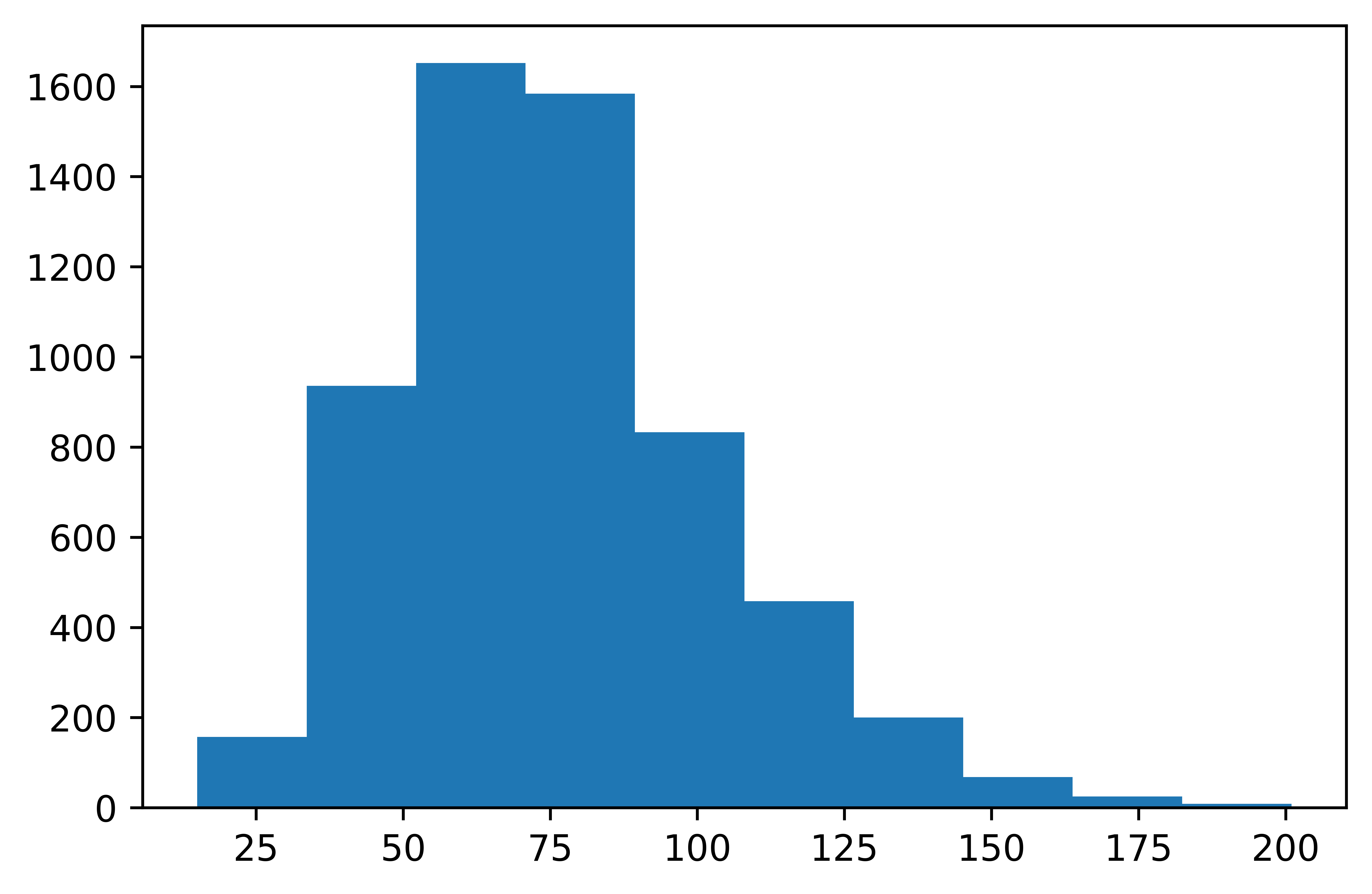}
  \caption{Query length distribution in character in the Vietnamese dataset.}
  \label{fig:vi_len_distribution}
\end{figure}

\begin{figure}
  \centering
\includegraphics[width=.9\linewidth]{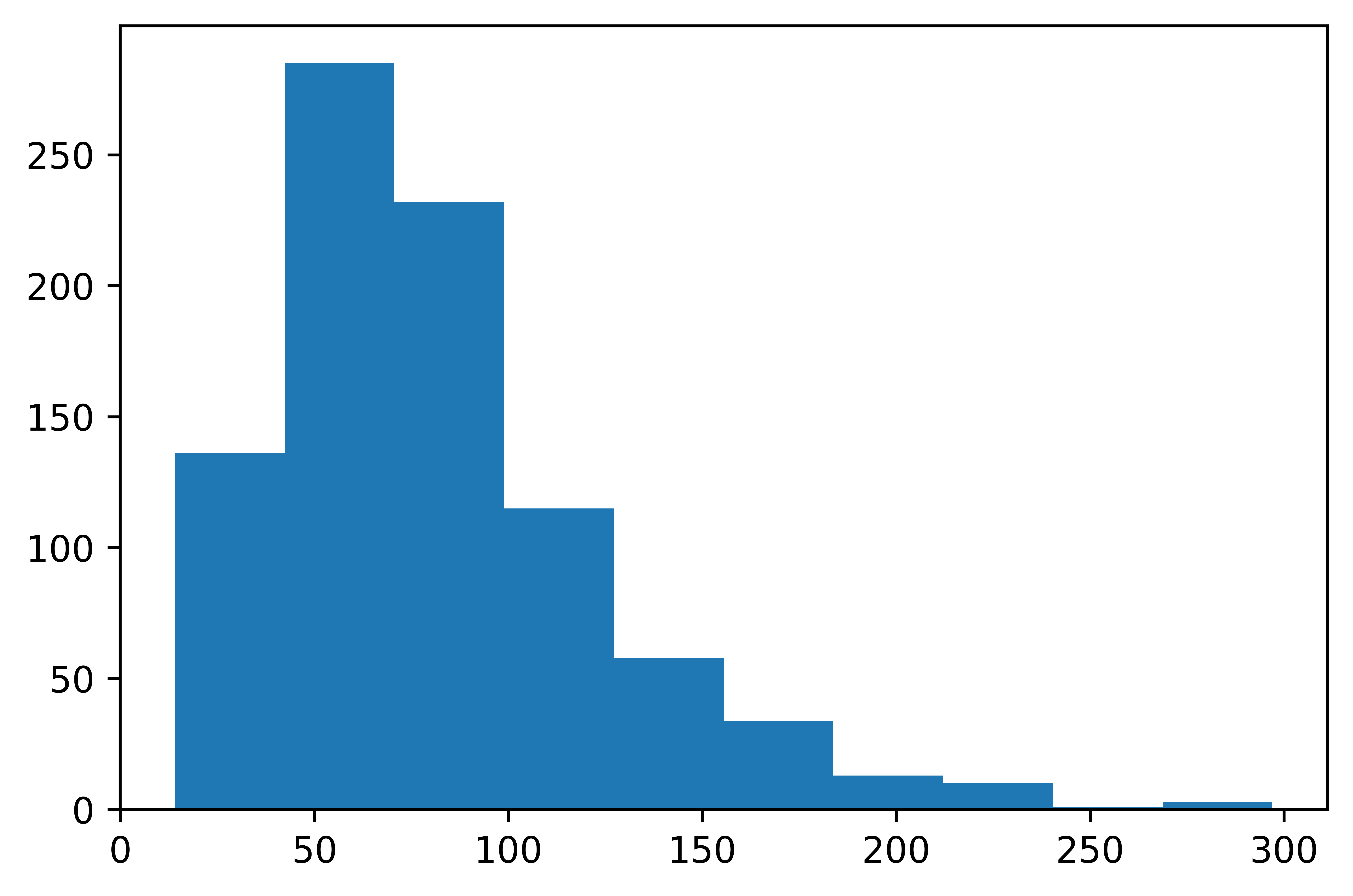}
  \caption{Query length distribution in character in the Japanese dataset.}
  \label{fig:jp_len_distribution}
\end{figure}

\begin{figure}
  \centering
\includegraphics[width=.9\linewidth]{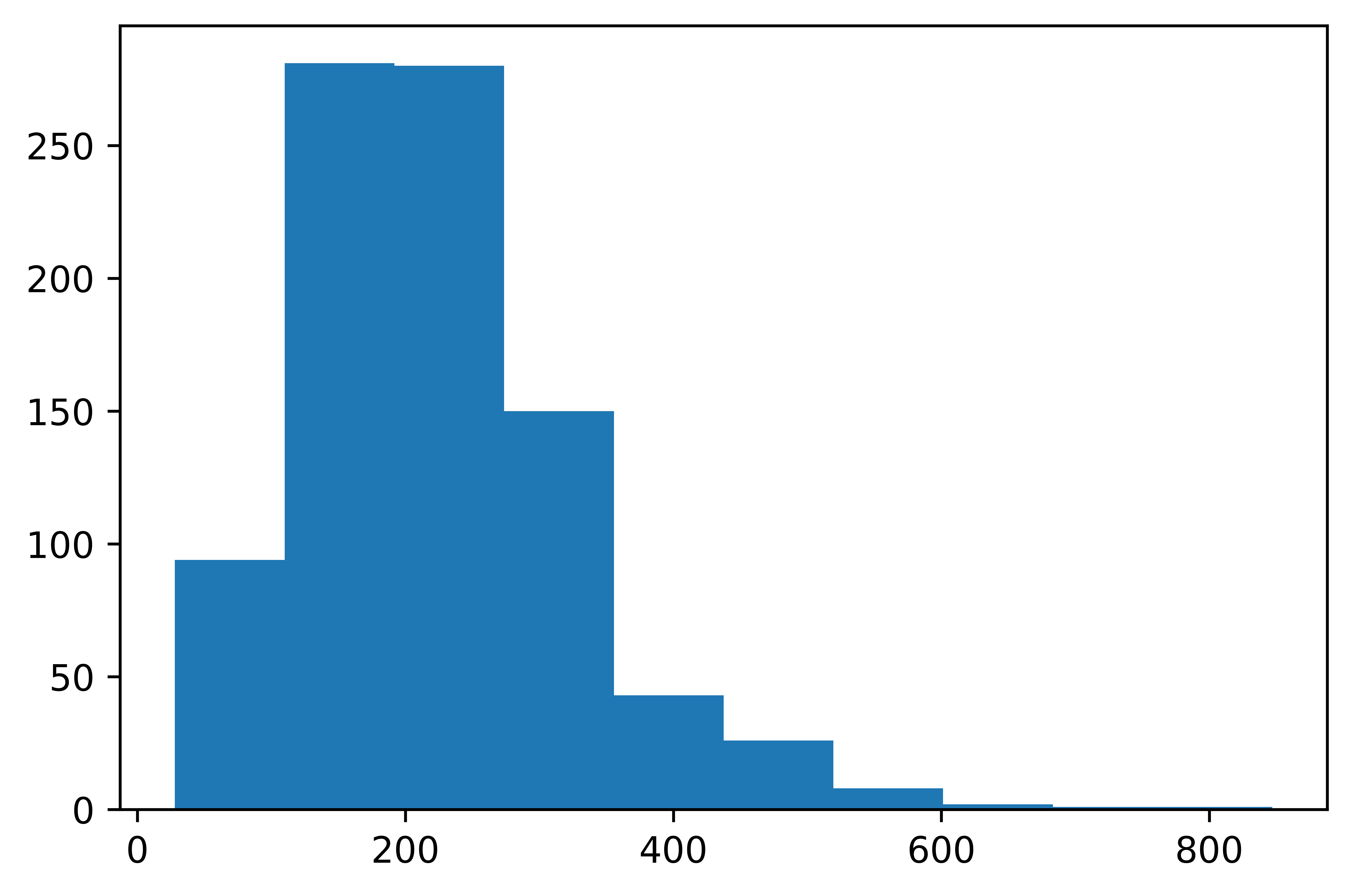}
  \caption{Query length distribution in character in the English dataset.}
  \label{fig:en_len_distribution}
\end{figure}

The English and Japanese data provided by COLIEE are of high quality. 
Though, the number of training samples is relatively small compared to the Vietnamese dataset, which is an interesting challenge for the deep learning approach.
The total number of samples to train the model is 806. The formal test set contains 81 samples.
The limitation in the amount of data makes it a practical situation to compare the performance of training-from-scratch models and pretrained models.

Figures~\ref{fig:vi_len_distribution},~\ref{fig:jp_len_distribution},~and~ \ref{fig:en_len_distribution} demonstrate the length distribution in characters of the queries in the Vietnamese, Japanese and English datasets respectively.
The Vietnamese dataset contains the largest number of queries and almost all of them are shorter than 200 characters.
The distribution suggests this dataset is suitable for training deep learning models from scratch.
The Japanese and English datasets contain not only fewer but also longer samples.
The longest sample is in the English dataset with more than 800 characters.
Datasets in multiple languages containing samples of varying lengths are useful for  analyzing the characteristics of different models.

\section{Retrieval Methods}
\label{sec:method}
\subsection{General Approach}

\begin{figure}
  \centering
\includegraphics[width=.9\linewidth]{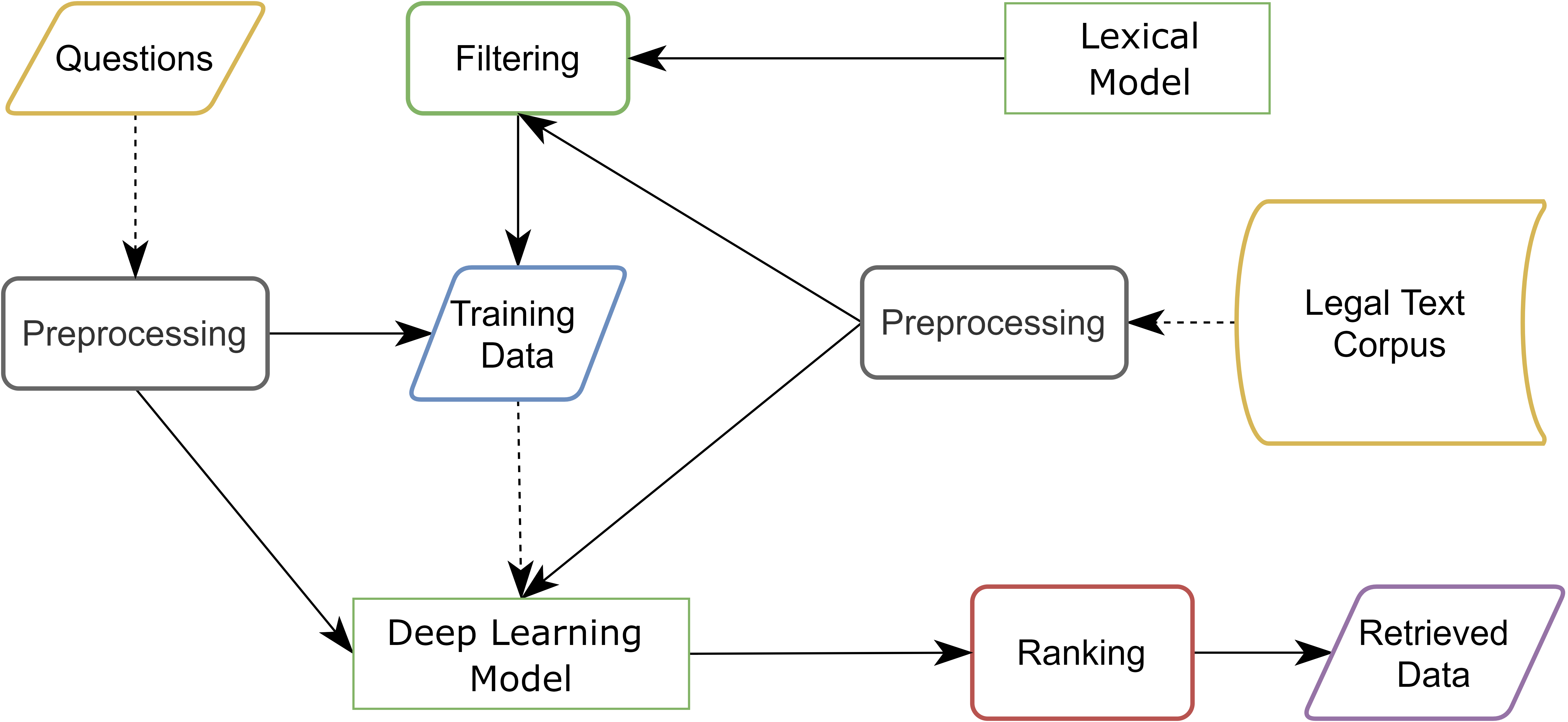}
  \caption{The pipeline of our proposed approach.}
  \label{fig:pipeline}
\end{figure}

The pipeline of our general approach is shown in Figure \ref{fig:pipeline}.
There are two phases in the process (\ie training and inference).
In the training phase, from the given question set and the legal text corpus, we preprocess the raw text into a proper form. 
To obtain the training data, we use the lexical model to filter out non-lexical-matched articles.
This process may also remove the relevant candidates from the data; however, this is the trade-off we have to take due to computational resource limitations.
After that, the deep learning model is trained by the negative sampling paradigm. 
In the inference phase, we combine the score from the trained model and the lexical score to rank the candidates to obtain the final relevant articles.

We propose two different architectures of deep neural networks with the general idea of \textit{divide-and-conquer}.
The first architecture uses convolutional networks without pretraining, which is named Attentive CNN, the second architecture leverages the power of the Transformer-based pretrained language model, which is named Paraformer.
Both architectures contain two main components, namely sentence encoder and paragraph encoder.
The sentence encoder is designed to encode legal sentences (\ie articles and queries) into vectors. 
The paragraph encoder aggregates the signal from the sentence encoder to obtain the final representation.
Finally, this representation is used to calculate the relevance between the query and the candidate article (paragraph).

To build the training data, we apply a negative sampling paradigm. 
With each query, along with the $P$ positive articles given by the ground truth, we sample $N$ negative articles from the corpus.
The model needs to predict the labels of each candidate in the set of $P+N$ articles.
In making training data for Attentive CNN, we combine both negative sampling using lexical matching and random negative sampling.
For Paraformer, we only sample negative candidates with high lexical overlapping with the query.

In the remaining part of this section, we introduce the detailed architecture of Attentive CNN and Paraformer and the way to train them to rank candidates given a query.
Considering that query has important information for the model to interpret the candidates in an appropriate aspect, in both designs, we inject the representation of the query as an input to construct the final article representation.

\subsection{Attentive CNN}
\subsubsection{Sentence Encoder}

\begin{figure}[ht]
  \centering
\includegraphics[width=.7\linewidth]{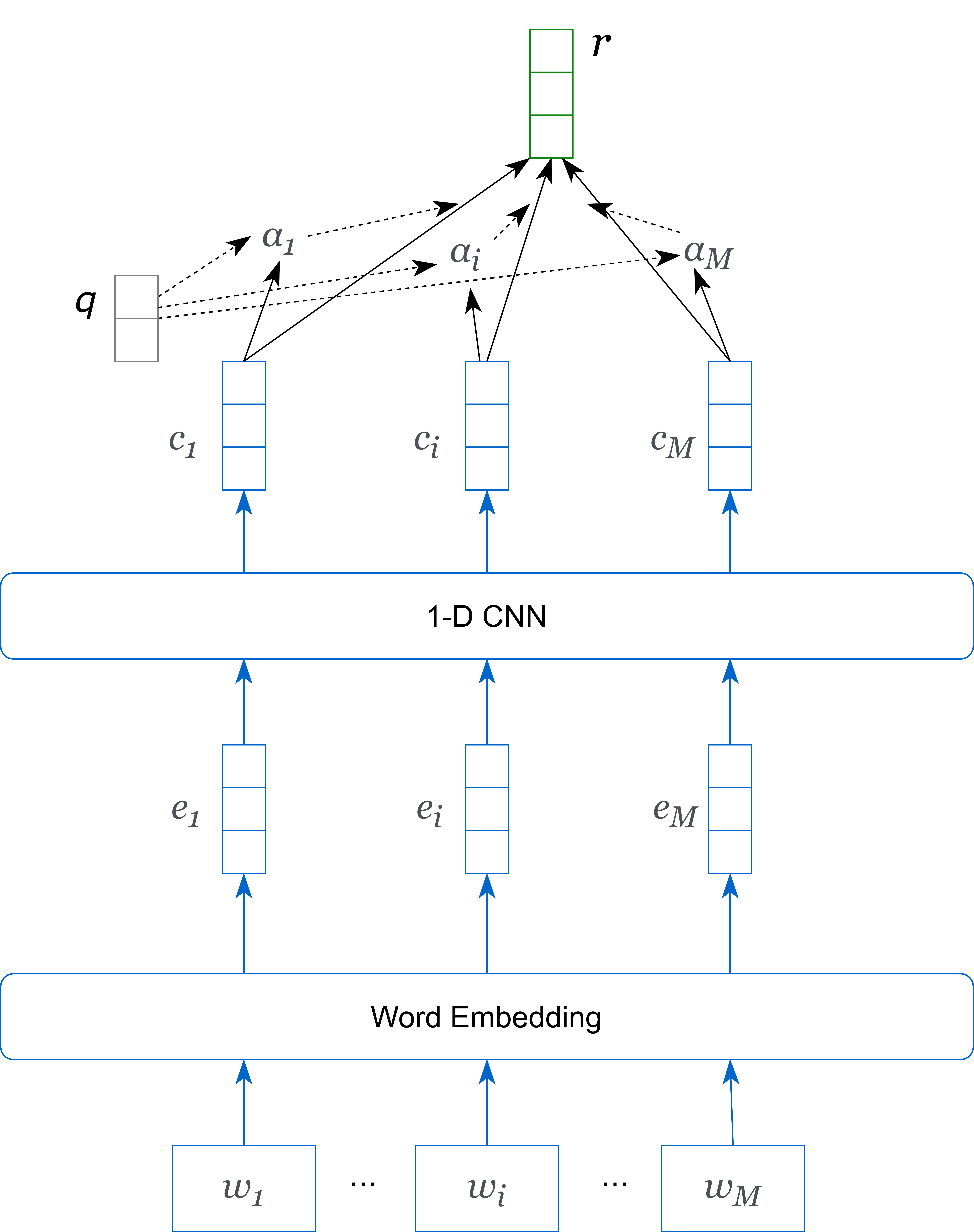}
  \caption{Sentence encoder component in Attentive CNN architecture.}
  \label{fig:cnn_sentenc}
\end{figure}

Figure~\ref{fig:cnn_sentenc} shows the architecture of our sentence encoder component in Attentive CNN. This component contains three layers: word embedding, convolution, and attention layers. With $M$ be the length of the input, word embedding is a mapping matrix from the index of the words $(w_1, w_2, ..., w_M)$ into corresponding vectors $(e_1, e_2, ..., e_M)$. The convolution layer aggregates the outputs of word embeddings to produce a more abstract vector $c_i$  for each position $i$ in the input considering the context formed by the surrounding words (\eg ``river \textit{bank}'' should be distinguished from ``financial \textit{bank}''). 

\highlight{
With $e_{(i-K):(i+K)}$ be the vector at the positions from $(i-K)$ to $(i+K)$, $F \in \mathbb{R}^{N_{f} \times(2 K+1) D}$ and $b_{t} \in \mathbb{R}^{N_{f}}$ be the kernel and the bias of the convolutional layer, $N_{f}$ be the number of filters, $2K + 1$ be the window size, $D$ be the vector dimension, the formula calculates the context $c_i$ of the word $i$ is as in Equation \ref{eq:ci}.
}

\begin{equation}
\label{eq:ci}
c_{i}=\operatorname{ReLU}\left(F \times e_{(i-K):(i+K)}\right)+b_{t}
\end{equation}

\highlight{
The attention layer is designed to calculate how important each word contributes to answering a given query. Let $q$ be the attention query vector, attention weight $a_i$ and normalized attention weight $\alpha_i$ of the word $i$ are calculated by Equations~\ref{eq:a_s}~and~\ref{eq:alpha_s} with $V$ and $v$ be the weight matrix and the bias value.
}

\begin{align}
    a_{i}&=q^{T} \tanh \left(V \times c_{i}+v\right)\label{eq:a_s}\\
    \alpha_{i}&=\frac{\exp \left(a_{i}\right)}{\sum_{j=1}^{M} \exp \left(a_{j}\right)}\label{eq:alpha_s}
\end{align}

The final representation vector $r$ is the weighted sum of $c_i$, as follows:

\begin{equation}
    r=\sum_{i=1}^{M} \alpha_{i} c_{i}
\end{equation}

\subsubsection{Paragraph Encoder}

\begin{figure}
  \centering
\includegraphics[width=.7\linewidth]{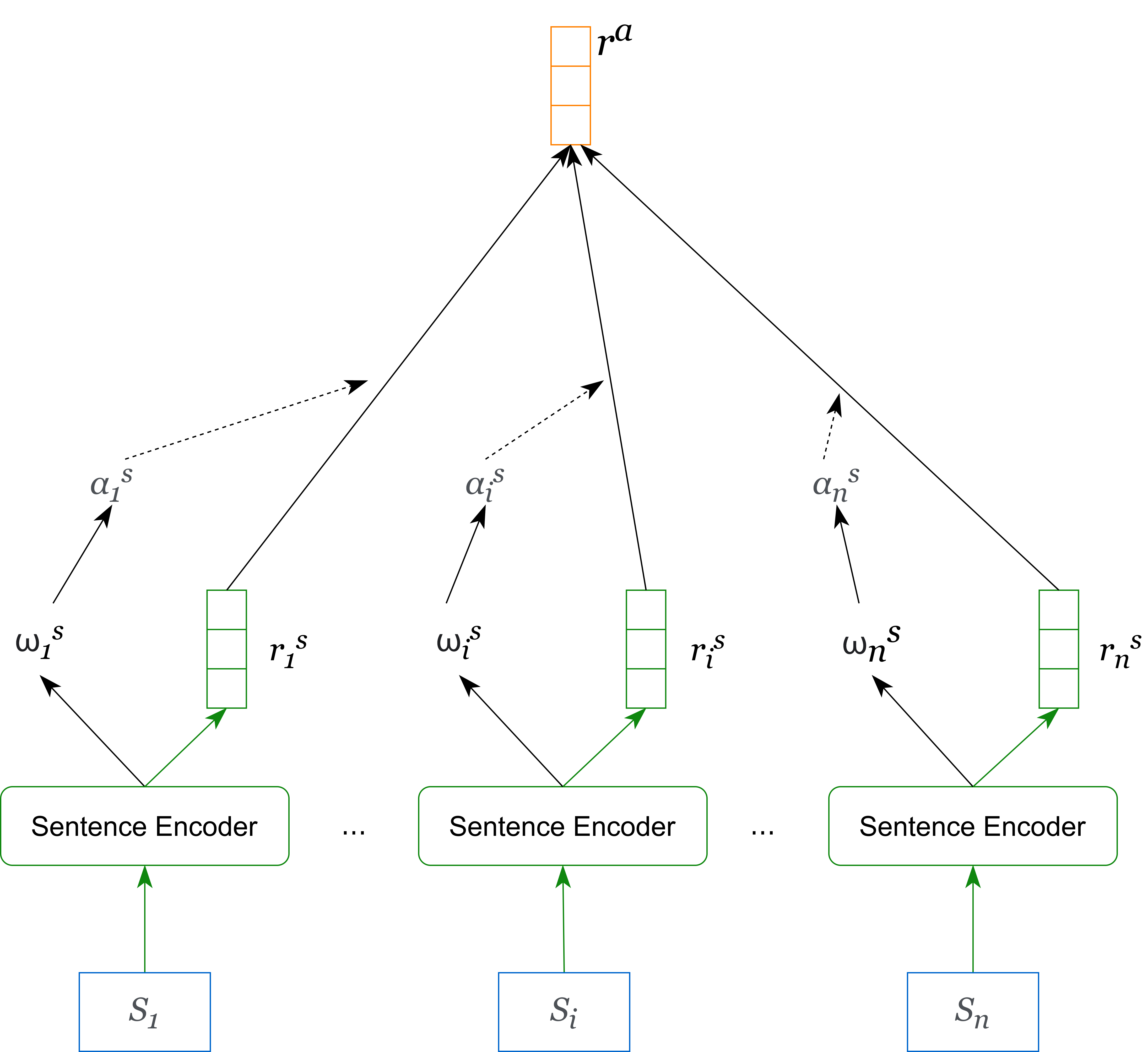}
  \caption{Paragraph encoder component in Attentive CNN architecture.}
  \label{fig:cnn_paraenc}
\end{figure}

An article in a legal document is often presented in a paragraph (\ie a set of sentences). 
We design a module called \textit{paragraph encoder} whose architecture is demonstrated in Figure~\ref{fig:cnn_paraenc}. 
This architecture shows the \textit{divide-and-conquer} paradigm idea as presented.
Instead of using a language model to directly encode an article, we encode each sentence of it and combine the signals via a global attention mechanism.

In designing this component, we have an important observation about the semantic contribution in a legal paragraph.
No single sentence represents the whole meaning of the paragraph and each sentence contributes an amount of semantics differently to the entire semantics.
We can recognize this phenomenon by reading the example given in Table~\ref{tab:example}.
Only several sentences in the highlighted parts contribute most to the necessary information to answer the query.
Other parts are not much relevant and may be used to answer other queries.
For that reason, we propose to apply sparsemax~\citep{martins2016softmax} to aggregate the signal from each sentence. 
If we use a softmax or an average function in this case, the required signal may be incomplete or diluted.

The representation vector $r^a$ of a paragraph is calculated by Equations~\ref{eq:omega_s},~\ref{eq:alpha_si},~and~\ref{eq:ra}.
Let $|s|$ be the number of words in the sentence $s$, the attention weight $\omega^{s}$ is the average value of the attention weights of the words belonging to that sentence as in Equation \ref{eq:omega_s}.

\begin{align}
\omega^{s}&=\frac{\sum_{i} a_{i}^{w}}{|s|}\label{eq:omega_s}
\end{align}

\highlight{
The normalized attention weight $\alpha_{j}^{s}$ and the final representation $r^{a}$ are calculated as in Equations \ref{eq:alpha_si} and \ref{eq:ra} with $N$ being the number of sentences in the paragraph, $\omega_{j}^{s}$ and $r_{j}^{s}$ be the original attention weight and the representation vector of the $j^{th}$ sentence in the paragraph. Sparsemax function ~\citep{martins2016softmax} produces the Euclidean projection of the input vector  $\omega_{j}^{s}$ onto the probability simplex.
}

\begin{align}
\alpha_{j}^{s}&=\operatorname{sparsemax}\left(\omega_{j}^{s}\right)\label{eq:alpha_si}\\
r^{a}&=\sum_{j=1}^{N} \alpha_{j}^{s} r_{j}^{s}\label{eq:ra}
\end{align}

With the proposed approach, the system learns to focus on the important parts and ignore other irrelevant ones. 
Besides, with the ability to highlight the important sentences in a lengthy article, the system can benefit the real user experience in its application.

\subsubsection{Model Training}

\begin{figure}
  \centering
\includegraphics[width=.55\linewidth]{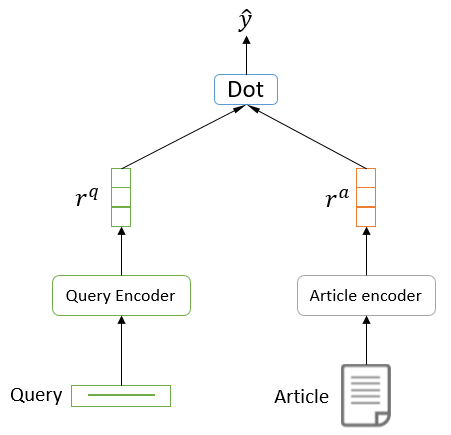}
  \caption{Training Attentive CNN as a similarity function.}
  \label{fig:cnn_similarity}
\end{figure}

We assign the components proposed above as backbones in our Attentive CNN architecture as demonstrated in Figure \ref{fig:cnn_similarity} and train them using the negative sampling paradigm.
In this approach, we encode the query and the article using the sentence encoder component and the paragraph encoder component to get corresponding representation vectors.
We then use dot product between the two vectors as the similarity score.
\highlight{
We normalize the similarity score as in Equation~\ref{eq:nomalize}. Given a query $q$, $\hat{y}_{i}^{+}$ is the probability that the article $i$ related to $q$, $\hat{y}_{i, j}^{-}$ is such probability that the article $j$ in the negative set of the article $i$ related to $q$, and $K$ is the number of articles in the sampled negative set.
}

\begin{equation}
\label{eq:nomalize}
p_{i}=\frac{\exp \left(\hat{y}_{i}^{+}\right)}{\exp \left(\hat{y}_{i}^{+}\right)+\sum_{j=1}^{K} \exp \left(\hat{y}_{i, j}^{-}\right)}
\end{equation}

\subsection{Paraformer}
\subsubsection{Sentence Encoder}

\begin{figure}[ht]
  \centering
\includegraphics[width=.6\linewidth]{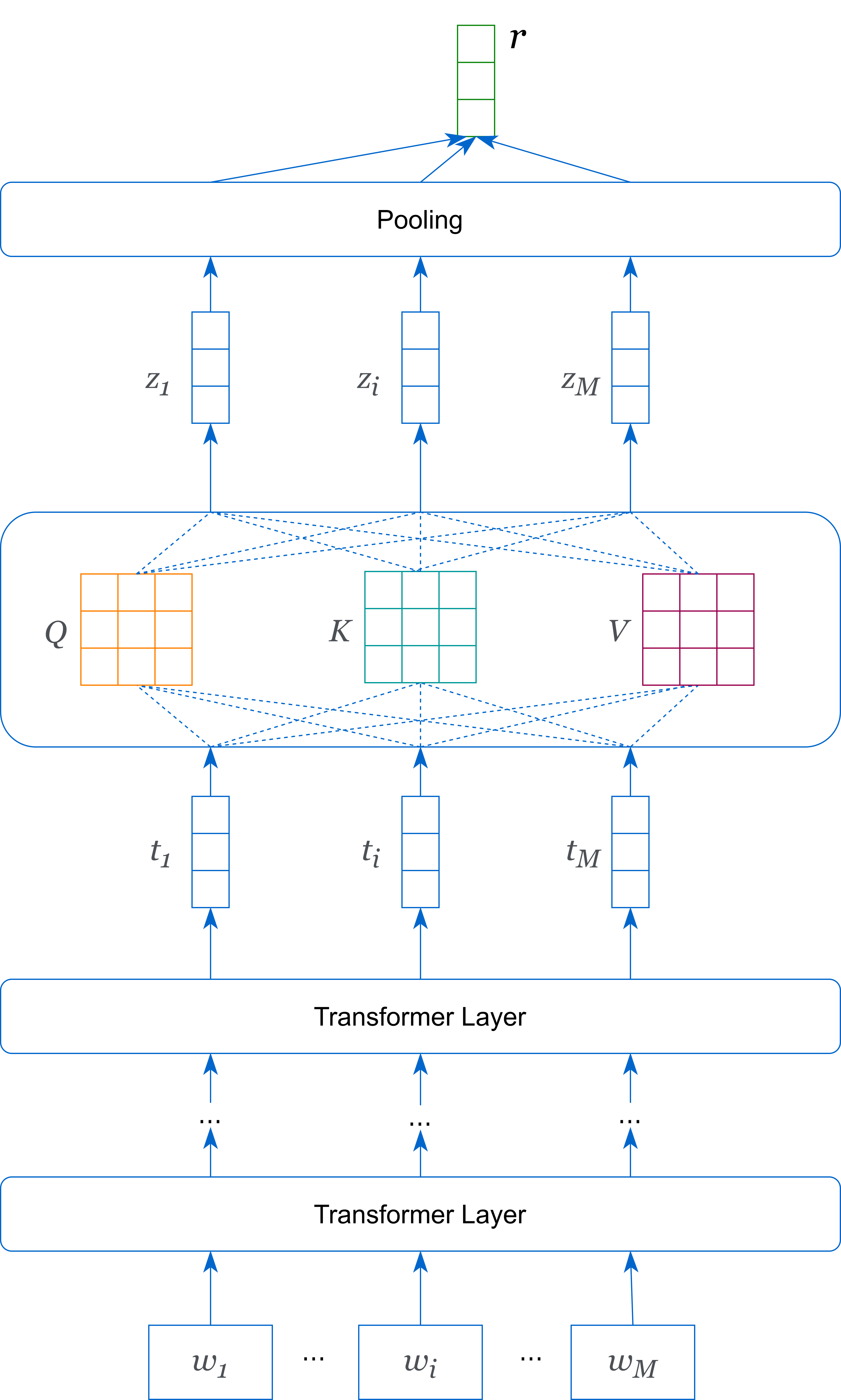}
  \caption{Sentence encoder component in Paraformer architecture.}
  \label{fig:paraformer_sentence_encoder_architecture}
\end{figure}

Attentive CNN's sentence encoder can work effectively with a sufficient amount of data \citep{kien-etal-2020-answering}.
However, like other training-from-scratch approaches, this component may struggle with problems with small amounts of data.
We confirm this issue in Section~\ref{sec:experiments}.
For the problem with limited data, this component shows severely reduced performance. 
For that reason, we propose to replace this component with a pretrained language model.
As in Figure~\ref{fig:paraformer_sentence_encoder_architecture}, the signal of an $M$-token input is transformed using the self-attention mechanism through the transformer layers. After that, the vectors in the final transformer layer are fed through a pooling layer to obtain a sentence-level representation vector.





\subsubsection{Paragraph Encoder}

\begin{figure}
  \centering
\includegraphics[width=.9\linewidth]{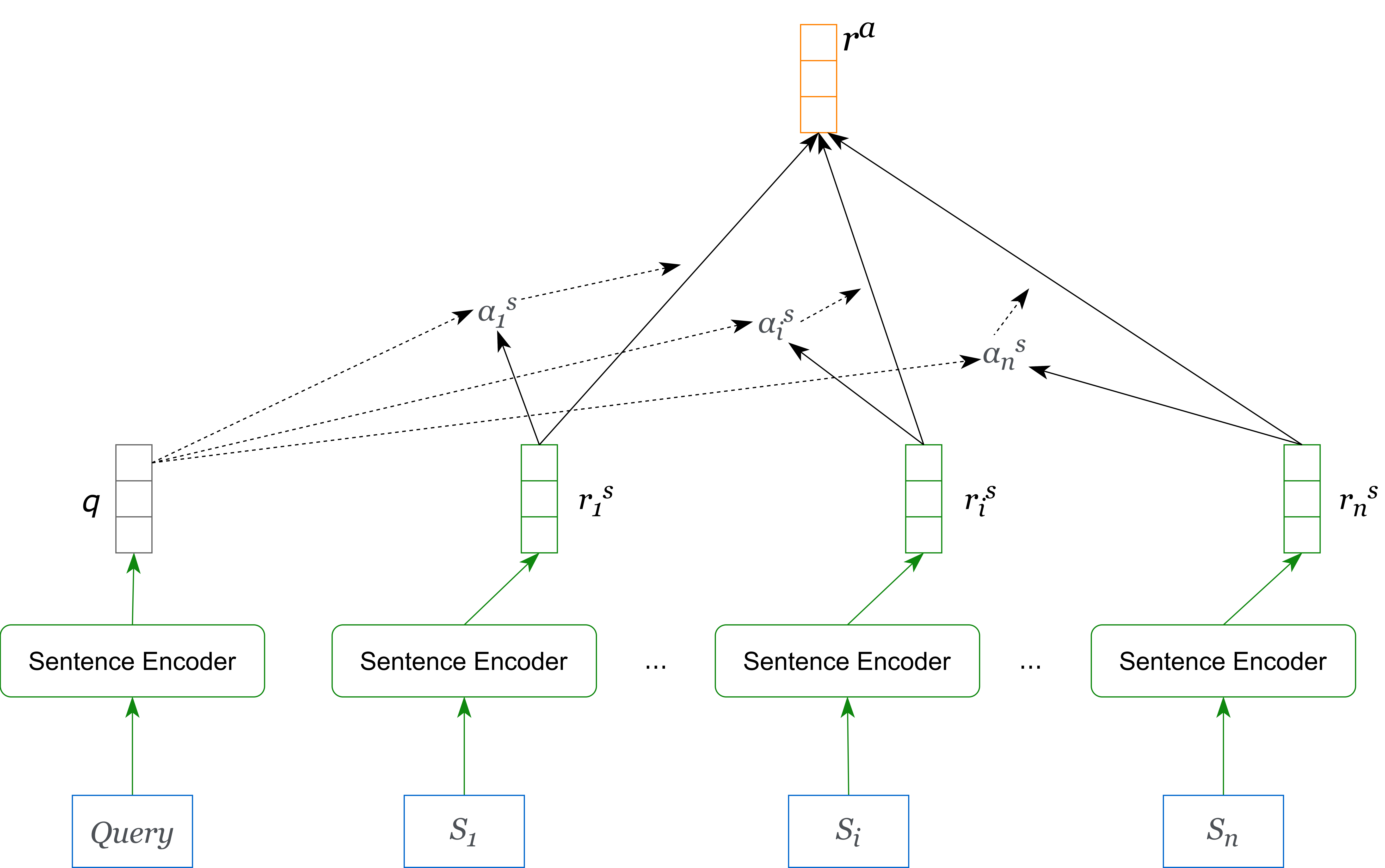}
  \caption{Paragraph encoder component in Paraformer architecture.}
  \label{fig:paraformer_paragraph_encoder_architecture}
\end{figure}

Unlike the paragraph encoder of Attentive CNN, the paragraph encoder of Paraformer incorporates query information with sentences in the article based on general attention, as in Figure~\ref{fig:paraformer_paragraph_encoder_architecture}.
We first produce sentence-level representations of the query ($q$), and $n$ sentences in an article ($r_1^s - r_n^s$) with the sentence encoder component.
\highlight{
Then, with general attention, the representation of an article for the given query is calculated by  Equations~\ref{eq:a_i_s},~\ref{eq:alpha_i_s},~and~\ref{eq:r_a}, with $A$ being the weight matrix, $b$ being the bias value.
}

\begin{align}
    a_i^s&=q^{T} \tanh \left(A \times r_i^s+b\right)\label{eq:a_i_s}\\
    \alpha_{i}^{s}&=\operatorname{sparsemax}\left(a_{i}^{s}\right)\label{eq:alpha_i_s}\\
    r^a&=\sum_{i=1}^{M} \alpha_i^s r_i^s\label{eq:r_a}
\end{align}

\subsubsection{Model Training}

As described in the design of this architecture, the sentence encoder is the unit component of the paragraph encoder.
In addition, this unit, which contains multi-head attention layers, is already pretrained with a large amount of data.
With Paraformer, we put one fully connected layer on top of the paragraph encoder and treat the whole model as a binary classifier.
We also use the cross-entropy loss in this approach.
Training this model is essentially updating the weights of global attention and finetuning the pretrained weights for a similarity prediction problem.
In the inference phase, we extract the logit value from the fully connected layer as the ranking score of this model.


\section{Experiments}
\label{sec:experiments}

\subsection{Experimental Settings}
The experiments are conducted with COLIEE's datasets and the Vietnamese dataset introduced in Section \ref{sec:dataset}. 
In the Vietnamese dataset, we used 90\% of the query set for training and validation, and the test set is 10\%. 
For English and Japanese, we use COLIEE 2021 data with the same train/test division as in the official competition.
We compare Attentive CNN, Paraformer and the vanilla XLM-RoBERTa, which is a strong multilingual pretrained baseline.
\highlight{
On the English dataset, we also experiment with BERT-PLI \citep{shao2020bert}, a very successful model for English legal retrieval of common law (Task 1, 2 of COLIEE 2019).
}

The Attentive CNN is trained from scratch, so it can perform in all three languages. 
The size of the vocabulary of this model is 31,450.
For the backbone of Paraformer's sentence encoder, among pretrained models provided by \citet{reimers2019sentence}, we choose \textit{paraphrase-xlm-r-multilingual-v1} for the multilingual version (including Japanese and Vietnamese), and \textit{paraphrase-mpnet-base-v2} for the English version.
The size of the vocabulary in the English version is 30,527 and in the multilingual version is 250,002.
Table \ref{tab:parameters_value_system2} and Table \ref{tab:parameters_value_paraformer} indicate the parameters of our two models (\ie Attentive CNN and Paraformer).
\highlight{
For BERT-PLI, we also finetune this model with case law entailment data as suggested by the authors before training the model on article retrieval data. 
Before conducting the experiment, we did not expect a model designed for the document level of case law to work well at the article level of statute law.}

\begin{table}
\caption{\label{tab:parameters_value_system2}
Value of parameters in Attentive CNN}
\begin{center}
\begin{tabular}{p{5cm} r}
\toprule  \textbf{Parameter} & \textbf{Value} \\
\midrule 
Size of Word Embedding layer & 512 \\
Number of CNN filter & 512 \\
Size of attention query vector & 200 \\
Dropout rate & 0.2 \\
\bottomrule
\end{tabular}
\end{center}
\end{table}

\begin{table}
\caption{\label{tab:parameters_value_paraformer}
Value of parameters in Paraformer}
\begin{center}
\begin{tabular}{p{5cm} r}
\toprule  \textbf{Parameter} & \textbf{Value} \\
\midrule 
Max Position Embeddings & 514\\
Hidden Size & 768 \\
Hidden Layers & 12 \\
Attention Heads & 12 \\
Dropout rate & 0.1 \\
\bottomrule
\end{tabular}
\end{center}
\end{table}

For all systems, we retrieve the articles in two stages: lexical matching and reranking.
In the lexical matching stage, for the Vietnamese dataset, because of the huge number of articles, we use ElasticSearch\footnote{https://www.elastic.co/} and for English and Japanese datasets, we use a lightweight python package Rank-BM25\footnote{https://pypi.org/project/rank-bm25/}.

\highlight{
In the reranking stage, we rank the articles using the final score calculated in  Equation \ref{eq:scoring_function}.
\begin{equation}
\label{eq:scoring_function}
S_{final}=\alpha \cdot S_{deep}+(1-\alpha) \cdot S_{lexical}
\end{equation}
where lexical score $S_{lexical}$ is obtained from the lexical matching system, and the semantic score $ S_{deep}$ is given by the deep learning model. 
$\alpha \in [0, 1]$, which can be tuned using hyperparameter tuning techniques, determines the weight between the two scores. 
}

We use the same metrics with COLIEE 2021, in which Macro-F2 at top 1 is the main metric to measure the performance of retrieval systems.
We also consider Precision and Recall scores for the analysis purpose.


\subsection{Experimental Results on COLIEE Datasets}
COLIEE datasets have been used by many research groups. 
This helps us better validate our methods and compare them with already presented systems.
We conduct the experiment in two phases.
At first, we compare different deep learning candidates' performances on the datasets without the support of BM25 (\ie $\alpha=1$).
After that, we apply a grid search optimization to our best candidate to know the highest performance our method can achieve.

\begin{table}
\centering
\caption{Performance of the systems without using the lexical score ($\alpha=1$)}\label{tab:coliee_res}
\begin{tabular}{|l|c|c|c|}
\hline
\textbf{Systems} & \textbf{Precision} & \textbf{Recall} & \textbf{F2}     \\ \hline
\multicolumn{4}{|c|}{English Dataset}                                     \\ \hline
\textbf{Paraformer}       & 0.3827             & 0.3450          & \textbf{0.3498} \\ \hline
XLM-RoBERTa      & 0.2099             & 0.1975          & 0.1989          \\ \hline
\highlight{BERT-PLI}         & \highlight{0.1728}             & \highlight{0.1543}          & \highlight{0.1564}          \\ \hline
Attentive CNN    & 0.0864             & 0.0864          & 0.0864          \\ \hline
\multicolumn{4}{|c|}{Japanese Dataset}                                    \\ \hline
\textbf{Paraformer }      & 0.3457             & 0.3148          & \textbf{0.3182} \\ \hline
XLM-RoBERTa      & 0.2940             & 0.3086          & 0.3086          \\ \hline
Attentive CNN    & 0.2593             & 0.2222          & 0.2263          \\ \hline
\end{tabular}
\end{table}

The first phase's results are shown in Table~\ref{tab:coliee_res}.
Paraformer achieves state-of-the-art results in both languages.
\highlight{BERT-PLI, a  model proposed for case law retrieval, surprised us with significantly better performance than Attentive CNN on the English dataset.
This can be explained by the ability of the deep learning models in transferring knowledge between similar data domains.
From this result, we can observe that pretrained models may be able to overcome situations in which data is not abundant.
}

Next, we tune the model to reach the optimal configurations in COLIEE 2021's formal dataset.
In the first phase, Paraformer achieves state-of-the-art results on the English dataset.
We choose this model as the deep learning component to combine with BM25 in the optimized reranking phase.
In this paper, the full table of grid-search can be found in Appendix \ref{app:grid_search}.



\begin{table}
\caption{Performance of Paraformer* compared with other competitors on COLIEE 2021's official test}
\label{tab:performance_coliee}
\center
\begin{tabular}{|l|c|c|c|}
\hline
\textbf{Run ID}                        & \textbf{Precision} & \textbf{Recall} & \textbf{F2} \\ \hline
\textbf{Paraformer*}                   & \textbf{0.7901}             & 0.7346          & \textbf{0.7407}       \\ \hline
OvGU     \citep{wehnert2021legal}              & 0.6749             & 0.7778          & 0.7302      \\ \hline
JNLP \citep{nguyen2021jnlp}   & 0.6000             & \textbf{0.8025}          & 0.7227      \\ \hline
UA   \citep{kim2022legal}                    & 0.7531             & 0.7037          & 0.7092      \\ \hline
TR  \citep{schilder2021pentapus}                      & 0.3333             & 0.6173          & 0.5226      \\ \hline
HUKB  \citep{masaharu2021bert}                      & 0.2901             & 0.6975          & 0.5224      \\ \hline
\end{tabular}

\end{table}

Table \ref{tab:performance_coliee} shows the performance of our final system (\ie \textit{Paraformer*}) compared to the state-of-the-art approaches from different teams in COLIEE 2021.
\textit{Paraformer*} obtains state-of-the-art performance on Precision and Macro-F2.
The best Recall performance belongs to the systems of \citet{nguyen2021jnlp} and \citet{wehnert2021legal}.
It could be room for future improvement.

\subsection{Experimental Results on Vietnamese Dataset}

Vietnamese dataset is larger than the COLIEE's datasets. 
Conducting an experiment on this dataset allows us to understand more about the behavior of the models.
In this dataset, we compare 4 candidates as follows:
\begin{itemize}

\item \textbf{BM25}: A well-known retrieval system using only the lexical features.

\item \textbf{XLM-RoBERTa}: Transformer-based model pretrained on a multilingual dataset in 100 languages \citep{conneau2019unsupervised} including English, Japanese and Vietnamese.

\item \textbf{Attentive CNN}: The convolutional neural network with the global attention mechanism.

\item \textbf{Paraformer}: Our novel proposed system taking advantage of the pretrained language model and the global attention.

\end{itemize}

\begin{table}
\centering
\caption{Experimental Results on Vietnamese Dataset on top-1 article.}\label{tab:vi_res}
\begin{tabular}{|l|c|c|c|}
\hline
\textbf{Systems} & \textbf{Precision} & \textbf{Recall} & \textbf{F2} \\ \hline
BM25             & 0.2395                                  & 0.1966                               & 0.2006                           \\ \hline
XLM-RoBERTa      & 0.2395                                  & 0.1966                               & 0.2006                           \\ \hline
Attentive CNN    & 0.5919                                  & 0.4660                               & 0.4774                           \\ \hline
Paraformer       & \textbf{0.5987}                                  & \textbf{0.4769}                               & \textbf{0.4882}                  \\ \hline
\end{tabular}
\end{table}

Table \ref{tab:vi_res} shows the experimental results on the Vietnamese dataset.
As we can see in the table, XLM-RoBERTa contributes no significant improvement compared to BM25 in Macro-F2 (0.2006).
Our Attentive CNN and Paraformer lead the ranking, Paraformer (0.4882) slightly outperforms Attentive CNN (0.4774) by about 1\%.
In our experiments, because of computation complexity, the number of articles filtered by lexical matching $N$ for Paraformer (from 10 to 150 articles) is significantly smaller than for the Attentive CNN (from 300 to 2,000 articles).
Curious about this difference, we further measure the performance on the top 20 articles retrieved by the two models, Attentive CNN achieves 0.2220 in Macro-F2@20 and 0.5849 in NDCG@20 while Paraformer achieves only 0.1839 and 0.4464, respectively.
This suggests that, for searching many results over a large search space, Attentive CNN might be a more suitable approach.

Despite being a pretrained model, XLM-RoBERTa performs badly in the Vietnamese dataset.
Analyzing the dataset, we see that the average length of Vietnamese legal sentences is significantly longer than English and Japanese sentences. 
In addition, concatenating the query and articles to construct the input for the system makes more burden on this model.
Even a powerful model can perform badly if they do not have full information for inference.
This strengthens the usefulness of the models proposed in this paper with the idea of \textit{divide-and-conquer}.

\subsection{Further Discussions}

\vspace{0.5cm}
\textbf{Impact of Content Length}

\begin{table}
\centering
\caption{Length in characters of Vietnamese, English and Japanese test sets.\label{tab:test_length}}
\begin{tabular}{|l|R{1.2cm}|R{1.2cm}|R{1.2cm}|R{1.2cm}|R{1.2cm}|R{1.2cm}|}
\hline
\multirow{2}{*}{\textbf{Dataset}} & \multicolumn{3}{c|}{\textbf{Query Length}} & \multicolumn{3}{c|}{\textbf{Article Length}} \\ \cline{2-7} 
                                  & \textbf{Min}         & \textbf{Max}      & \textbf{Avg.}        & \textbf{Min}       & \textbf{Max}        & \textbf{Avg.}      \\ \hline
Vietnamese                        & 20          & 182      & 78             & 53        & 252,955    & 10,941              \\ \hline
English                           & 60          & 379      & 214            & 203       & 1,891       & 742             \\ \hline
Japanese                          & 21          & 219      & 90             & 58        & 550        & 224             \\ \hline
\end{tabular}
\end{table}
Table \ref{tab:test_length} indicates the length in characters of the Vietnamese, English and Japanese testing sets. 
Note that, since each model has a different way of tokenizing input sentences, in this paper, we use the number of characters as a common unit to measure the length of samples.
In the Vietnamese dataset, the length of articles varies greatly, the longest article is about 250K characters, the shortest article is 53 characters. 
The pretrained models have a limit of 514 tokens.
This creates a significant challenge for vanilla XLM-RoBERTa with the approach of treating an entire article as a sentence.
Looking at Table \ref{tab:coliee_res}, \ref{tab:vi_res} and \ref{tab:test_length}, we have the observation that XLM-RoBERTa may obtain poor results with too lengthy articles.

\label{subsec:different_length}
\begin{figure}[ht]
  \centering
\includegraphics[width=.9\linewidth]{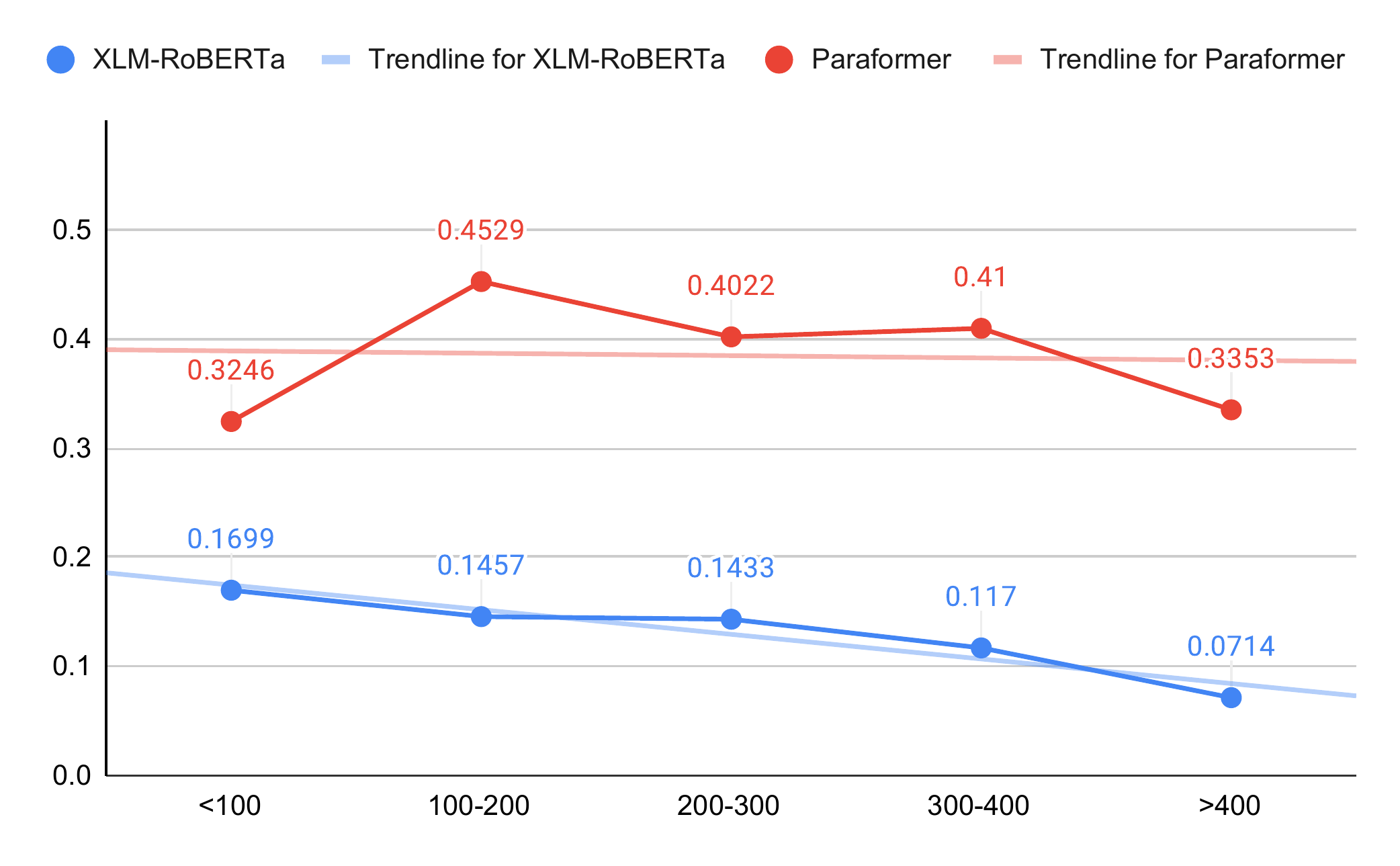}
  \caption{Performance of XLM-RoBERTa and Paraformer when working on different lengths of queries. \highlight{The x-axis represents the length of the query chunk in characters, the y-axis represents the performance of the models in Macro F2.}}
  \label{fig:diffrent_length}
\end{figure}


Figure \ref{fig:diffrent_length} shows the performance of the XLM-RoBERTa and Paraformer along with their trendlines on different chunks of query length in the English dataset.
It can be seen that the longer the query, the worse the performance of both models.
However, we can see that Paraformer is the winner in all chunks and its trendline reduces slower.

\vspace{0.5cm}
\noindent \textbf{Global Attention Visualization}

\begin{figure}[ht]
  \centering
\includegraphics[width=.95\linewidth]{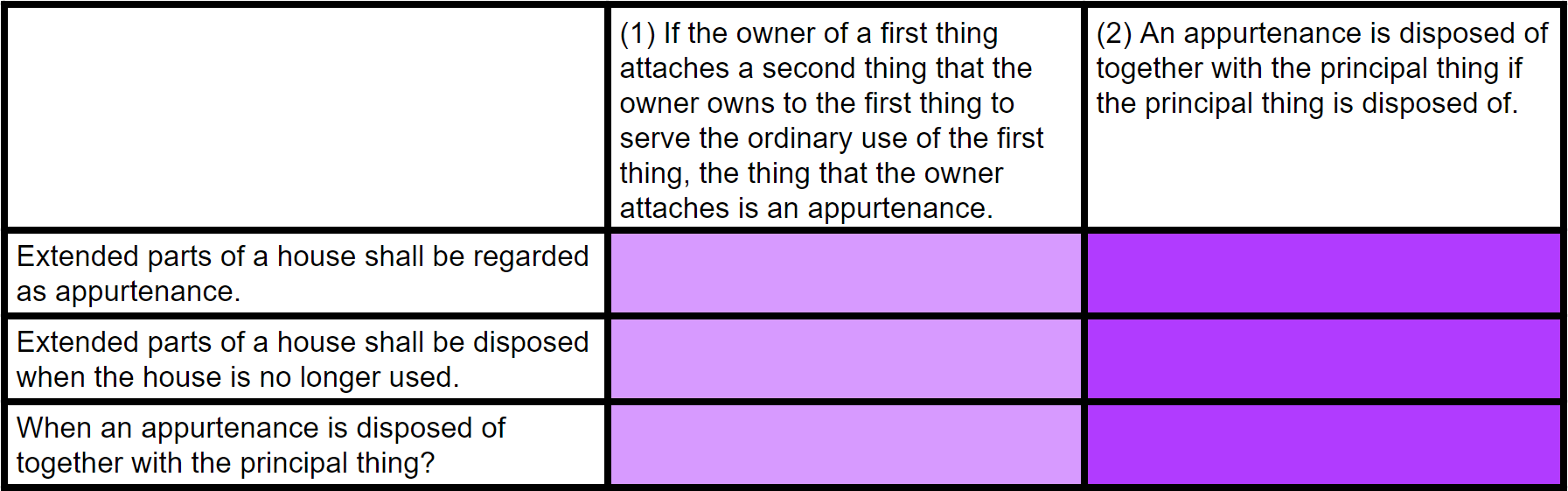}
  \caption{Weight visualization of Attentive CNN for the example in Section \ref{sec:intro}. The more important the content, the darker the color.}
  \label{fig:attentive_cnn_weight_visualization}
\end{figure}
\begin{figure}[ht]
  \centering
\includegraphics[width=.95\linewidth]{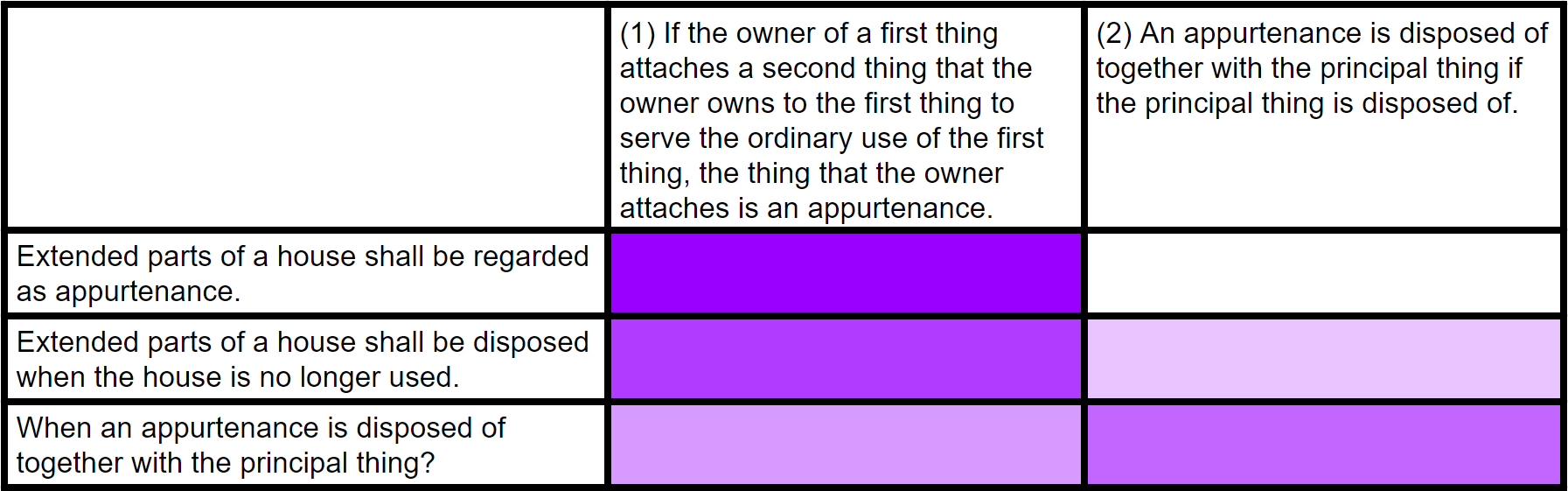}
  \caption{Weight visualization of Paraformer for the example in Section \ref{sec:intro}. The more important the content, the darker the color. }
  \label{fig:paraformer_weight_visualization}
\end{figure}

\highlight{
Although sharing a common \textit{divide-and-conquer} idea with Attentive CNN, the architecture of Paraformer allows us to represent the relevance between the queries and the articles more flexibly.
After being trained, while the Attentive CNN generates only one article representation regardless of the query, Paraformer's paragraph encoder allows us to derive information about the relevance between queries and each sentence in an article through its attention weights. Figure \ref{fig:attentive_cnn_weight_visualization} and \ref{fig:paraformer_weight_visualization} demonstrate the attention weights of Attentive CNN and Paraformer for the same example mentioned in Section \ref{sec:intro}.
}
\highlight{
As we can see in the figure, Paraformer focuses differently on the contents of Article 87 depending on the given query while Attentive CNN produces the same attention weights for all queries.
This also opens up interesting research directions in explainable AI where we can debug what information the models are paying attention to instead of accepting their results as black-box output.
} 

\section{Conclusions}

In this paper, we investigate and solve the problem of information retrieval for the legal domain by using deep learning models with the attention mechanism to represent the query and article for the ranking purpose.
The general idea of our approach, \textit{divide-and-conquer}, is to break down articles to represent them individually and then combine them back using global attention.
We propose two new architectures named Attentive CNN and Paraformer based on this idea.
In our experiment, we demonstrate the effectiveness of this method compared to strong baselines in reliable legal datasets in three different languages, \ie, English, Japanese, and Vietnamese.
We also analyze the strengths and weaknesses of each model with each specific data condition for a clear insight in designing the models for this problem.
In addition, our large Vietnamese dataset for this problem enables us to perform detailed analysis as well as to contribute to the research community.
In future work, we intend to extend this work by introducing more legal domain-specific pretrained methods for this architecture.

\section*{Acknowledgements}
This work was supported by JSPS Kakenhi Grant Number 20K20406. The research also was supported in part by the Asian Office of Aerospace R\&D(AOARD), AirForce Office of Scientific Research (Grant no. FA2386-19-1-4041). The work would not be complete without valuable data from COLIEE.

\clearpage
\begin{appendices}

\section{Data Examples}

\begin{table}[ht]
\caption{\label{tab:example}
A sample in the Vietnamese dataset with highlighted parts}
\selectlanguage{vietnamese}
\begin{center}
\begin{tabular}{|l|p{80mm}|}
\hline Question & Con riêng có được hưởng di sản thừa kế của người cha đã mất khi không để lại di chúc không? \\
\hline Answer & Article 651 from the Code of Civil law of Vietnam (2015). \\
\hline Article content & Điều 651. \newline \textbf{\textit{Người thừa kế theo pháp luật}} \newline \textbf{\textit{1. Những người thừa kế theo pháp luật được quy định theo thứ tự sau đây:}} \newline \textbf{\textit{a) Hàng thừa kế thứ nhất gồm: vợ, chồng, cha đẻ, mẹ đẻ, cha nuôi, mẹ nuôi, con đẻ, con nuôi của người chết;}} \newline  b) Hàng thừa kế thứ hai gồm: ông nội, bà nội, ông ngoại, bà ngoại, anh ruột, chị ruột, em ruột của người chết; cháu ruột của người chết mà người chết là ông nội, bà nội, ông ngoại, bà ngoại; \newline c) Hàng thừa kế thứ ba gồm: cụ nội, cụ ngoại của người chết; bác ruột, chú ruột, cậu ruột, cô ruột, dì ruột của người chết; cháu ruột của người chết mà người chết là bác ruột, chú ruột, cậu ruột, cô ruột, dì ruột; chắt ruột của người chết mà người chết là cụ nội, cụ ngoại. \newline 2. Những người thừa kế cùng hàng được hưởng phần di sản bằng nhau. \newline 3. Những người ở hàng thừa kế sau chỉ được hưởng thừa kế, nếu không còn ai ở hàng thừa kế trước do đã chết, không có quyền hưởng di sản, bị truất quyền hưởng di sản hoặc từ chối nhận di sản. \\

\hline
\end{tabular}
\end{center}
\end{table}

\selectlanguage{english}

\begin{table}[ht]
\caption{\label{tab:jp_example}
A sample in the Japanese dataset}
\begin{center}
\begin{tabular}{|l|p{80mm}|}
\hline Question & \begin{CJK*}{UTF8}{min}未成年者がした売買契約は、親権者の同意を得ないでした場合であっても、その契約が日常生活に関するものであるときは、取り消すことができない。\end{CJK*} \\
\hline Answer & Article 5 from Japanese Civil Code. \\
\hline Article content & \begin{CJK*}{UTF8}{min}第五条　未成年者が法律行為をするには、その法定代理人の同意を得なければならない。ただし、単に権利を得、又は義務を免れる法律行為については、この限りでない。\end{CJK*} \newline
\begin{CJK*}{UTF8}{min}２　前項の規定に反する法律行為は、取り消すことができる。\end{CJK*} \newline
\begin{CJK*}{UTF8}{min}３　第一項の規定にかかわらず、法定代理人が目的を定めて処分を許した財産は、その目的の範囲内において、未成年者が自由に処分することができる。目的を定めないで処分を許した財産を処分するときも、同様とする。\end{CJK*} \\
\hline
\end{tabular}
\end{center}
\end{table}

\begin{table}
\caption{\label{tab:en_example}
A sample in the English dataset}
\begin{center}
\begin{tabular}{|l|p{80mm}|}
\hline Question & A contract of guarantee concluded by a person under curatorship may not be rescinded in cases the consent of the curator is obtained. \\
\hline Answer & Article 13 from Japanese Civil Code. \\
\hline Article content & Article 13 \newline
(1) A person under curatorship must obtain the consent of the curator in order to perform any of the following acts;provided, however, that this does not apply to an act provided for in the proviso of Article 9: \newline
(i) receiving or using any property producing civil fruit;\newline
(ii) borrowing money or guaranteeing an obligation;\newline
(iii) performing an act with the purpose of acquiring or losing any right regarding immovables or other significant property;\newline
(iv) suing any procedural act;\newline
(v) giving a gift, reaching a settlement, or entering into an arbitration agreement (meaning an arbitration agreement as provided in Article 2, paragraph (1) of the Arbitration Act (Act No. 138 of 2003));\newline
(vi) accepting or renouncing a succession or dividing an estate;\newline
(vii) refusing an offer of a gift, renouncing a legacy, accepting an offer of gift with burden, or accepting a legacy with burden;\newline
(viii) constructing a new building, renovating, expanding, or undertaking major repairs;\newline
(ix) granting a lease for a term that exceeds the period set forth in Article 602; or\newline
(x) performing any of the acts set forth in the preceding items as a legal representative of a person with qualified legal capacity (meaning a minor, adult ward, or person under curatorship or a person under assistance who is subject to a decision as referred to in Article 17, paragraph (1); the same applies hereinafter).\newline
(2) At the request of a person as referred to in the main clause of Article 11\newline
or the curator or curator's supervisor, the family court may decide that the person under curatorship must also obtain the consent of the curator before performing an act other than those set forth in each of the items of the preceding paragraph;provided, however, that this does not apply to an act provided for in the proviso to Article 9.\newline
(3) If the curator does not consent to an act for which the person under curatorship must obtain the curator's consent even though it is unlikely to prejudice the interests of the person under curatorship, the family court may grant permission that operates in lieu of the curator's consent at the request of the person under curatorship.\newline
(4) An act for which the person under curatorship must obtain the curator's consent is voidable if the person performs it without obtaining the curator's consent or a permission that operates in lieu of it.. \\
\hline
\end{tabular}
\end{center}
\end{table}

\clearpage
\section{Grid Search Table for Tuning Paraformer*}
\label{app:grid_search}
\begin{longtable}{| c |c |c |c |c |c |c |} 
\hline
\multirow{2}{*}{\textbf{$\alpha$}} & \multicolumn{3}{c|}{\textbf{Validation}}           & \multicolumn{3}{c|}{\textbf{Test}}                 \\ \cline{2-7} 
                                & \textbf{P} & \textbf{R} & \textbf{F2} & \textbf{P} & \textbf{R} & \textbf{F2} \\ \hline
\multicolumn{7}{|c|}{\textbf{Top\_BM25=10}}                                                                                               \\ \hline
0.1                             & 0.5077             & 0.4692          & 0.4735      & 0.6790             & 0.6481          & 0.6516      \\ \hline
0.2                             & 0.5231             & 0.4846          & 0.4889      & 0.6790             & 0.6481          & 0.6516      \\ \hline
0.3                             & 0.5231             & 0.4846          & 0.4889      & 0.6790             & 0.6481          & 0.6516      \\ \hline
0.4                             & 0.5846             & 0.5462          & 0.5504      & 0.6914             & 0.6543          & 0.6584      \\ \hline
0.5                             & 0.6000             & 0.5615          & 0.5658      & 0.6914             & 0.6543          & 0.6584      \\ \hline
0.6                             & 0.6308             & 0.5923          & 0.5966      & 0.7160             & 0.6790          & 0.6831      \\ \hline
0.7                             & 0.6462             & 0.6000          & 0.6051      & 0.7531             & 0.7099          & 0.7147      \\ \hline
0.8                             & 0.6154             & 0.5692          & 0.5744      & 0.7654             & 0.7160          & 0.7215      \\ \hline
0.9                             & 0.6154             & 0.5615          & 0.5675      & 0.7901             & 0.7346          & 0.7407      \\ \hline
1.0                             & 0.5231             & 0.4462          & 0.4547      & 0.3827             & 0.3457          & 0.3498      \\ \hline
\multicolumn{7}{|c|}{\textbf{Top\_BM25=20}}                                                                                               \\ \hline
0.1                             & 0.5077             & 0.4692          & 0.4735      & 0.6790             & 0.6481          & 0.6516      \\ \hline
0.2                             & 0.5231             & 0.4846          & 0.4889      & 0.6790             & 0.6481          & 0.6516      \\ \hline
0.3                             & 0.5231             & 0.4846          & 0.4889      & 0.6790             & 0.6481          & 0.6516      \\ \hline
0.4                             & 0.5846             & 0.5462          & 0.5504      & 0.6914             & 0.6543          & 0.6584      \\ \hline
0.5                             & 0.6000             & 0.5615          & 0.5658      & 0.6914             & 0.6543          & 0.6584      \\ \hline
0.6                             & 0.6308             & 0.5923          & 0.5966      & 0.7160             & 0.6790          & 0.6831      \\ \hline
0.7                             & 0.6462             & 0.6000          & 0.6051      & 0.7654             & 0.7222          & 0.7270      \\ \hline
0.8                             & 0.6154             & 0.5692          & 0.5744      & 0.7778             & 0.7284          & 0.7339      \\ \hline
0.9                             & 0.5846             & 0.5385          & 0.5436      & 0.7654             & 0.7160          & 0.7215      \\ \hline
1.0                             & 0.4154             & 0.3462          & 0.3538      & 0.2840             & 0.2593          & 0.2620      \\ \hline
\multicolumn{7}{|c|}{\textbf{Top\_BM25=30}}                                                                                               \\ \hline
0.1                             & 0.5077             & 0.4692          & 0.4735      & 0.6790             & 0.6481          & 0.6516      \\ \hline
0.2                             & 0.5231             & 0.4846          & 0.4889      & 0.6790             & 0.6481          & 0.6516      \\ \hline
0.3                             & 0.5231             & 0.4846          & 0.4889      & 0.6790             & 0.6481          & 0.6516      \\ \hline
0.4                             & 0.5846             & 0.5462          & 0.5504      & 0.6914             & 0.6543          & 0.6584      \\ \hline
0.5                             & 0.6000             & 0.5615          & 0.5658      & 0.6914             & 0.6543          & 0.6584      \\ \hline
0.6                             & 0.6308             & 0.5923          & 0.5966      & 0.7160             & 0.6790          & 0.6831      \\ \hline
0.7                             & 0.6462             & 0.6000          & 0.6051      & 0.7654             & 0.7222          & 0.7270      \\ \hline
0.8                             & 0.6154             & 0.5692          & 0.5744      & 0.7778             & 0.7284          & 0.7339      \\ \hline
0.9                             & 0.5692             & 0.5308          & 0.5350      & 0.7654             & 0.7160          & 0.7215      \\ \hline
1.0                             & 0.3077             & 0.2538          & 0.2598      & 0.1605             & 0.1543          & 0.1550      \\ \hline
\multicolumn{7}{|c|}{\textbf{Top\_BM25=40}}                                                                                               \\ \hline
0.1                             & 0.5077             & 0.4692          & 0.4735      & 0.6790             & 0.6481          & 0.6516      \\ \hline
0.2                             & 0.5231             & 0.4846          & 0.4889      & 0.6790             & 0.6481          & 0.6516      \\ \hline
0.3                             & 0.5231             & 0.4846          & 0.4889      & 0.6790             & 0.6481          & 0.6516      \\ \hline
0.4                             & 0.5846             & 0.5462          & 0.5504      & 0.6914             & 0.6543          & 0.6584      \\ \hline
0.5                             & 0.6000             & 0.5615          & 0.5658      & 0.6914             & 0.6543          & 0.6584      \\ \hline
0.6                             & 0.6308             & 0.5923          & 0.5966      & 0.7160             & 0.6790          & 0.6831      \\ \hline
0.7                             & 0.6462             & 0.6000          & 0.6051      & 0.7654             & 0.7222          & 0.7270      \\ \hline
0.8                             & 0.6154             & 0.5692          & 0.5744      & 0.7778             & 0.7284          & 0.7339      \\ \hline
0.9                             & 0.5692             & 0.5205          & 0.5256      & 0.7778             & 0.7284          & 0.7339      \\ \hline
1.0                             & 0.2308             & 0.1821          & 0.1871      & 0.1481             & 0.1420          & 0.1427      \\ \hline
\multicolumn{7}{|c|}{\textbf{Top\_BM25=50}}                                                                                               \\ \hline
0.1                             & 0.5077             & 0.4692          & 0.4735      & 0.6790             & 0.6481          & 0.6516      \\ \hline
0.2                             & 0.5231             & 0.4846          & 0.4889      & 0.6790             & 0.6481          & 0.6516      \\ \hline
0.3                             & 0.5231             & 0.4846          & 0.4889      & 0.6790             & 0.6481          & 0.6516      \\ \hline
0.4                             & 0.5846             & 0.5462          & 0.5504      & 0.6914             & 0.6543          & 0.6584      \\ \hline
0.5                             & 0.6000             & 0.5615          & 0.5658      & 0.6914             & 0.6543          & 0.6584      \\ \hline
0.6                             & 0.6308             & 0.5923          & 0.5966      & 0.7160             & 0.6790          & 0.6831      \\ \hline
0.7                             & 0.6462             & 0.6000          & 0.6051      & 0.7654             & 0.7222          & 0.7270      \\ \hline
0.8                             & 0.6154             & 0.5692          & 0.5744      & 0.7778             & 0.7284          & 0.7339      \\ \hline
0.9                             & 0.5692             & 0.5205          & 0.5256      & 0.7778             & 0.7284          & 0.7339      \\ \hline
1.0                             & 0.2462             & 0.1974          & 0.2025      & 0.1481             & 0.1420          & 0.1427      \\ \hline
\multicolumn{7}{|c|}{\textbf{Top\_BM25=60}}                                                                                               \\ \hline
0.1                             & 0.5077             & 0.4692          & 0.4735      & 0.6790             & 0.6481          & 0.6516      \\ \hline
0.2                             & 0.5231             & 0.4846          & 0.4889      & 0.6790             & 0.6481          & 0.6516      \\ \hline
0.3                             & 0.5231             & 0.4846          & 0.4889      & 0.6790             & 0.6481          & 0.6516      \\ \hline
0.4                             & 0.5846             & 0.5462          & 0.5504      & 0.6914             & 0.6543          & 0.6584      \\ \hline
0.5                             & 0.6000             & 0.5615          & 0.5658      & 0.6914             & 0.6543          & 0.6584      \\ \hline
0.6                             & 0.6308             & 0.5923          & 0.5966      & 0.7160             & 0.6790          & 0.6831      \\ \hline
0.7                             & 0.6462             & 0.6000          & 0.6051      & 0.7654             & 0.7222          & 0.7270      \\ \hline
0.8                             & 0.6154             & 0.5692          & 0.5744      & 0.7654             & 0.7160          & 0.7215      \\ \hline
0.9                             & 0.5692             & 0.5205          & 0.5256      & 0.7778             & 0.7284          & 0.7339      \\ \hline
1.0                             & 0.2308             & 0.1821          & 0.1871      & 0.1358             & 0.1296          & 0.1303      \\ \hline
\multicolumn{7}{|c|}{\textbf{Top\_BM25=70}}                                                                                               \\ \hline
0.1                             & 0.5077             & 0.4692          & 0.4735      & 0.6790             & 0.6481          & 0.6516      \\ \hline
0.2                             & 0.5231             & 0.4846          & 0.4889      & 0.6790             & 0.6481          & 0.6516      \\ \hline
0.3                             & 0.5231             & 0.4846          & 0.4889      & 0.6790             & 0.6481          & 0.6516      \\ \hline
0.4                             & 0.5846             & 0.5462          & 0.5504      & 0.6914             & 0.6543          & 0.6584      \\ \hline
0.5                             & 0.6000             & 0.5615          & 0.5658      & 0.6914             & 0.6543          & 0.6584      \\ \hline
0.6                             & 0.6308             & 0.5923          & 0.5966      & 0.7160             & 0.6790          & 0.6831      \\ \hline
0.7                             & 0.6462             & 0.6000          & 0.6051      & 0.7654             & 0.7222          & 0.7270      \\ \hline
0.8                             & 0.6154             & 0.5692          & 0.5744      & 0.7654             & 0.7160          & 0.7215      \\ \hline
0.9                             & 0.5692             & 0.5205          & 0.5256      & 0.7778             & 0.7284          & 0.7339      \\ \hline
1.0                             & 0.2154             & 0.1846          & 0.1880      & 0.1358             & 0.1296          & 0.1303      \\ \hline
\multicolumn{7}{|c|}{\textbf{Top\_BM25=80}}                                                                                               \\ \hline
0.1                             & 0.5077             & 0.4692          & 0.4735      & 0.6790             & 0.6481          & 0.6516      \\ \hline
0.2                             & 0.5231             & 0.4846          & 0.4889      & 0.6790             & 0.6481          & 0.6516      \\ \hline
0.3                             & 0.5231             & 0.4846          & 0.4889      & 0.6790             & 0.6481          & 0.6516      \\ \hline
0.4                             & 0.5846             & 0.5462          & 0.5504      & 0.6914             & 0.6543          & 0.6584      \\ \hline
0.5                             & 0.6000             & 0.5615          & 0.5658      & 0.6914             & 0.6543          & 0.6584      \\ \hline
0.6                             & 0.6308             & 0.5923          & 0.5966      & 0.7160             & 0.6790          & 0.6831      \\ \hline
0.7                             & 0.6462             & 0.6000          & 0.6051      & 0.7654             & 0.7222          & 0.7270      \\ \hline
0.8                             & 0.6154             & 0.5692          & 0.5744      & 0.7654             & 0.7160          & 0.7215      \\ \hline
0.9                             & 0.5692             & 0.5205          & 0.5256      & 0.7778             & 0.7284          & 0.7339      \\ \hline
1.0                             & 0.2000             & 0.1692          & 0.1726      & 0.1111             & 0.1049          & 0.1056      \\ \hline
\multicolumn{7}{|c|}{\textbf{Top\_BM25=90}}                                                                                               \\ \hline
0.1                             & 0.5077             & 0.4692          & 0.4735      & 0.6790             & 0.6481          & 0.6516      \\ \hline
0.2                             & 0.5231             & 0.4846          & 0.4889      & 0.6790             & 0.6481          & 0.6516      \\ \hline
0.3                             & 0.5231             & 0.4846          & 0.4889      & 0.6790             & 0.6481          & 0.6516      \\ \hline
0.4                             & 0.5846             & 0.5462          & 0.5504      & 0.6914             & 0.6543          & 0.6584      \\ \hline
0.5                             & 0.6000             & 0.5615          & 0.5658      & 0.6914             & 0.6543          & 0.6584      \\ \hline
0.6                             & 0.6308             & 0.5923          & 0.5966      & 0.7160             & 0.6790          & 0.6831      \\ \hline
0.7                             & 0.6462             & 0.6000          & 0.6051      & 0.7654             & 0.7222          & 0.7270      \\ \hline
0.8                             & 0.6154             & 0.5692          & 0.5744      & 0.7654             & 0.7160          & 0.7215      \\ \hline
0.9                             & 0.5692             & 0.5205          & 0.5256      & 0.7778             & 0.7284          & 0.7339      \\ \hline
1.0                             & 0.1538             & 0.1308          & 0.1333      & 0.1111             & 0.1049          & 0.1056      \\ \hline
\multicolumn{7}{|c|}{\textbf{Top\_BM25=100}}                                                                                              \\ \hline
0.1                             & 0.5077             & 0.4692          & 0.4735      & 0.6790             & 0.6481          & 0.6516      \\ \hline
0.2                             & 0.5231             & 0.4846          & 0.4889      & 0.6790             & 0.6481          & 0.6516      \\ \hline
0.3                             & 0.5231             & 0.4846          & 0.4889      & 0.6790             & 0.6481          & 0.6516      \\ \hline
0.4                             & 0.5846             & 0.5462          & 0.5504      & 0.6914             & 0.6543          & 0.6584      \\ \hline
0.5                             & 0.6000             & 0.5615          & 0.5658      & 0.6914             & 0.6543          & 0.6584      \\ \hline
0.6                             & 0.6308             & 0.5923          & 0.5966      & 0.7160             & 0.6790          & 0.6831      \\ \hline
0.7                             & 0.6462             & 0.6000          & 0.6051      & 0.7654             & 0.7222          & 0.7270      \\ \hline
0.8                             & 0.6154             & 0.5692          & 0.5744      & 0.7654             & 0.7160          & 0.7215      \\ \hline
0.9                             & 0.5692             & 0.5205          & 0.5256      & 0.7654             & 0.7160          & 0.7215      \\ \hline
1.0                             & 0.1385             & 0.1231          & 0.1248      & 0.0988             & 0.0926          & 0.0933      \\ \hline
\multicolumn{7}{|c|}{\textbf{Top\_BM25=110}}                                                                                              \\ \hline
0.1                             & 0.5077             & 0.4692          & 0.4735      & 0.6790             & 0.6481          & 0.6516      \\ \hline
0.2                             & 0.5231             & 0.4846          & 0.4889      & 0.6790             & 0.6481          & 0.6516      \\ \hline
0.3                             & 0.5231             & 0.4846          & 0.4889      & 0.6790             & 0.6481          & 0.6516      \\ \hline
0.4                             & 0.5846             & 0.5462          & 0.5504      & 0.6914             & 0.6543          & 0.6584      \\ \hline
0.5                             & 0.6000             & 0.5615          & 0.5658      & 0.6914             & 0.6543          & 0.6584      \\ \hline
0.6                             & 0.6308             & 0.5923          & 0.5966      & 0.7160             & 0.6790          & 0.6831      \\ \hline
0.7                             & 0.6462             & 0.6000          & 0.6051      & 0.7654             & 0.7222          & 0.7270      \\ \hline
0.8                             & 0.6154             & 0.5692          & 0.5744      & 0.7654             & 0.7160          & 0.7215      \\ \hline
0.9                             & 0.5692             & 0.5205          & 0.5256      & 0.7654             & 0.7160          & 0.7215      \\ \hline
1.0                             & 0.1385             & 0.1231          & 0.1248      & 0.0741             & 0.0679          & 0.0686      \\ \hline
\multicolumn{7}{|c|}{\textbf{Top\_BM25=120}}                                                                                              \\ \hline
0.1                             & 0.5077             & 0.4692          & 0.4735      & 0.6790             & 0.6481          & 0.6516      \\ \hline
0.2                             & 0.5231             & 0.4846          & 0.4889      & 0.6790             & 0.6481          & 0.6516      \\ \hline
0.3                             & 0.5231             & 0.4846          & 0.4889      & 0.6790             & 0.6481          & 0.6516      \\ \hline
0.4                             & 0.5846             & 0.5462          & 0.5504      & 0.6914             & 0.6543          & 0.6584      \\ \hline
0.5                             & 0.6000             & 0.5615          & 0.5658      & 0.6914             & 0.6543          & 0.6584      \\ \hline
0.6                             & 0.6308             & 0.5923          & 0.5966      & 0.7160             & 0.6790          & 0.6831      \\ \hline
0.7                             & 0.6462             & 0.6000          & 0.6051      & 0.7654             & 0.7222          & 0.7270      \\ \hline
0.8                             & 0.6154             & 0.5692          & 0.5744      & 0.7654             & 0.7160          & 0.7215      \\ \hline
0.9                             & 0.5692             & 0.5205          & 0.5256      & 0.7654             & 0.7160          & 0.7215      \\ \hline
1.0                             & 0.1231             & 0.1154          & 0.1162      & 0.0741             & 0.0679          & 0.0686      \\ \hline
\multicolumn{7}{|c|}{\textbf{Top\_BM25=130}}                                                                                              \\ \hline
0.1                             & 0.5077             & 0.4692          & 0.4735      & 0.6790             & 0.6481          & 0.6516      \\ \hline
0.2                             & 0.5231             & 0.4846          & 0.4889      & 0.6790             & 0.6481          & 0.6516      \\ \hline
0.3                             & 0.5231             & 0.4846          & 0.4889      & 0.6790             & 0.6481          & 0.6516      \\ \hline
0.4                             & 0.5846             & 0.5462          & 0.5504      & 0.6914             & 0.6543          & 0.6584      \\ \hline
0.5                             & 0.6000             & 0.5615          & 0.5658      & 0.6914             & 0.6543          & 0.6584      \\ \hline
0.6                             & 0.6308             & 0.5923          & 0.5966      & 0.7160             & 0.6790          & 0.6831      \\ \hline
0.7                             & 0.6462             & 0.6000          & 0.6051      & 0.7654             & 0.7222          & 0.7270      \\ \hline
0.8                             & 0.6154             & 0.5692          & 0.5744      & 0.7654             & 0.7160          & 0.7215      \\ \hline
0.9                             & 0.5538             & 0.5154          & 0.5197      & 0.7654             & 0.7160          & 0.7215      \\ \hline
1.0                             & 0.1231             & 0.1154          & 0.1162      & 0.0741             & 0.0679          & 0.0686      \\ \hline
\multicolumn{7}{|c|}{\textbf{Top\_BM25=140}}                                                                                              \\ \hline
0.1                             & 0.5077             & 0.4692          & 0.4735      & 0.6790             & 0.6481          & 0.6516      \\ \hline
0.2                             & 0.5231             & 0.4846          & 0.4889      & 0.6790             & 0.6481          & 0.6516      \\ \hline
0.3                             & 0.5231             & 0.4846          & 0.4889      & 0.6790             & 0.6481          & 0.6516      \\ \hline
0.4                             & 0.5846             & 0.5462          & 0.5504      & 0.6914             & 0.6543          & 0.6584      \\ \hline
0.5                             & 0.6000             & 0.5615          & 0.5658      & 0.6914             & 0.6543          & 0.6584      \\ \hline
0.6                             & 0.6308             & 0.5923          & 0.5966      & 0.7160             & 0.6790          & 0.6831      \\ \hline
0.7                             & 0.6462             & 0.6000          & 0.6051      & 0.7654             & 0.7222          & 0.7270      \\ \hline
0.8                             & 0.6154             & 0.5692          & 0.5744      & 0.7654             & 0.7160          & 0.7215      \\ \hline
0.9                             & 0.5538             & 0.5154          & 0.5197      & 0.7654             & 0.7160          & 0.7215      \\ \hline
1.0                             & 0.1231             & 0.1154          & 0.1162      & 0.0741             & 0.0679          & 0.0686      \\ \hline
\multicolumn{7}{|c|}{\textbf{Top\_BM25=150}}                                                                                              \\ \hline
0.1                             & 0.5077             & 0.4692          & 0.4735      & 0.6790             & 0.6481          & 0.6516      \\ \hline
0.2                             & 0.5231             & 0.4846          & 0.4889      & 0.6790             & 0.6481          & 0.6516      \\ \hline
0.3                             & 0.5231             & 0.4846          & 0.4889      & 0.6790             & 0.6481          & 0.6516      \\ \hline
0.4                             & 0.5846             & 0.5462          & 0.5504      & 0.6914             & 0.6543          & 0.6584      \\ \hline
0.5                             & 0.6000             & 0.5615          & 0.5658      & 0.6914             & 0.6543          & 0.6584      \\ \hline
0.6                             & 0.6308             & 0.5923          & 0.5966      & 0.7160             & 0.6790          & 0.6831      \\ \hline
0.7                             & 0.6462             & 0.6000          & 0.6051      & 0.7654             & 0.7222          & 0.7270      \\ \hline
0.8                             & 0.6154             & 0.5692          & 0.5744      & 0.7654             & 0.7160          & 0.7215      \\ \hline
0.9                             & 0.5538             & 0.5154          & 0.5197      & 0.7654             & 0.7160          & 0.7215      \\ \hline
1.0                             & 0.1231             & 0.1154          & 0.1162      & 0.0741             & 0.0679          & 0.0686      \\ \hline
\end{longtable}

\end{appendices}

\clearpage
\bibliography{sn-bibliography}


\end{document}